\documentclass[aps,reprint,prb,showpacs,showkeys,preprintnumbers,amsmath,amssymb,citeautoscript,floatfix,superscriptaddress]{revtex4-1}
\usepackage{graphicx}
\usepackage{pstricks}
\usepackage{hyperref}
\usepackage[mathlines]{lineno}
\usepackage{mathtools}
\usepackage{subfigure}

\usepackage{dcolumn}
\usepackage{bm}

\usepackage{mathptmx}

\newcommand{\beginsupplement}{%
        \setcounter{table}{0}
        \renewcommand{\thetable}{S\arabic{table}}%
        \setcounter{figure}{0}
        \renewcommand{\thefigure}{S\arabic{figure}}%
     }

\newcounter{saveeqn}

\begin{document}

\title{Free-electron effects on optical absorption  of hybrid perovskite CH$_3$NH$_3$PbI$_3$ from first principles}

\author{Joshua Leveillee}
\affiliation{Department of Materials Science and Engineering, University of Illinois at Urbana-Champaign, Urbana, IL 61801, USA}
\author{Andr\'e Schleife}
\email{schleife@illinois.edu}
\affiliation{Department of Materials Science and Engineering, University of Illinois at Urbana-Champaign, Urbana, IL 61801, USA}
\affiliation{Frederick Seitz Materials Research Laboratory, University of Illinois at Urbana-Champaign, Urbana, IL 61801, USA}
\affiliation{National Center for Supercomputing Applications, University of Illinois at Urbana-Champaign, Urbana, IL 61801, USA}

\date{\today}

\begin{abstract}
Hybrid organic-inorganic perovskites, such as methyl-ammonium lead tri-iodide (MAPbI$_3$), are interesting candidates for efficient absorber materials in next-generation solar cells, partly due to an unusual combination of low exciton binding energy and strong optical absorption.
Excitonic effects in this material have been subject to debate both for experiment and theory, indicating a need for better understanding of the screening mechanisms that act upon the electron-hole interaction.
Here we use cutting-edge first-principles theoretical spectroscopy, based on density-functional and many-body perturbation theory, to study atomic geometries, electronic structure, and optical properties of three MAPbI$_3$ polymorphs and find good agreement with earlier results and experiment.
We then study the influence of free electrons on the electron-hole interaction and show that this explains consistently smaller exciton binding energies, compared to those in the material without free electrons.
Interestingly, we also find that the absorption line shape strongly resembles that of the spectrum without free electrons up to high free electron concentrations.
We explain this unexpected behavior by formation of Mahan excitons that dominate the absorption edge, making it robust against free-electron induced changes observed in other semiconductors.
\end{abstract}

\maketitle

\section{Introduction}

Hybrid organic-inorganic perovskites have seen unprecedented development over the last years, largely motivated by their potential as highly efficient absorber materials for next-generation solar cells.
Research on these materials for optoelectronic applications originates as far back as the 1990s;
hybrid perovskites were initially reported as dye-sensitizers in TiO$_2$ scaffolds in 2006 \cite{Green:2017,adams_TiO2_scaffold}.
Since their development as thin-film meso-superstructured photovoltaics in 2012 \cite{Lee:2012}, their photo-conversion efficiency has risen to over 22.7\,$\%$ \cite{Ishihara:1994,Saliba:2016,green_solar_2015,Zhou:2016}.
The most commonly studied material in this context is methyl-ammonium lead tri-iodide MAPbI$_3$, with MA=CH$_3$NH$_3$, owing to cheap solution synthesis and high performance metrics.
Besides photovoltaic applications, MAPbI$_3$ and its stoichiometric counterparts MA(Pb,Sn)(I,Br,Cl)$_3$ have shown promise in quantum dot fluorescence \cite{Deng:2018}, light-emitting diodes \cite{Zhang:2017}, and catalysis for water splitting \cite{Bin:2016}.

In addition to interesting applications, the combination of organic and heavy-metal constituents renders MAPbI$_3$ an ideal candidate to study fundamental phenomena.
One example is the strong spin-orbit interaction due to heavy atoms, that heavily reduces the band gap and dominates band dispersion near the conduction-band minimum\cite{even_importance_2013,Mosconi:2016}.
Another example, critically influencing whether a material is a good candidate for a photovoltaic absorber, is the electron-hole interaction:
If it is strong in a material, strongly bound excitonic states appear near the absorption onset.
These are associated with strong optical absorption that is beneficial for harvesting light using as little absorber material as possible.
At the same time, strong electron-hole interaction renders separation of electron-hole pairs challenging, which is detrimental in a photovoltaic device \cite{Gregg:2003}.
Interestingly, MAPbI$_3$ balances between low exciton binding energy and large optical absorption across the visible spectrum.
This facilitates efficient generation of electron-hole pairs that can be thermally separated and is beneficial for photo-current generation \cite{Herz:2016}.

This interesting balance triggered numerous studies, aimed at a better understanding of excitonic effects in MAPbI$_3$.
Experimental results for exciton binding energies range from as high as 62 meV to as low as 2 meV \cite{Herz:2016}, however, a few patterns emerge:
First, the line shape of the absorption edge has been reported to be comparable to that of GaAs, with no clear excitonic peak and a binding energy potentially under 10 meV at room temperature \cite{Chen:2018}. 
Second, a reduction of the exciton binding energy is observed when going from the low-temperature (LT) orthorhombic phase to the room-temperature (RT) tetragonal phase.
Sestu \emph{et al.}\ measured 34 meV (LT) to 29 meV (RT) \cite{Sestu_2015}, Galkowski \emph{et al.}\ measured between 14 and 25 meV (LT) to 12 meV (RT) \cite{Galkowski_2016}, and Yang \emph{et al.}\ measured 16 meV (LT) to between 5 and 12 meV (RT) \cite{Yang:2015}.
However, there are also examples for studies where
RT exciton binding energies exceed LT binding energies in others \cite{D'Innocenzo:2014,Yang:2015,Sestu_2015,Savenije:2014,Lin:2014,Miyata:2015,Sun_2014,Galkowski_2016,Yang:2017}.

Further insight into this variability comes from four-wave mixing spectroscopy, to disentangle exciton binding energies of intrinsic and defect-bound excitons \cite{March:2016}.
These results indicate that intrinsic excitons have an LT binding energy of 13 meV, whereas values for defect-bound excitons average around 29 meV, linking the variability to different defect concentrations lest uncharacterized.
In particular, while exciton binding energies in pure MAPbI$_3$ are consistently lower than 35 meV for LT and RT phases, the addition of small amounts of chlorine into MAPbI$_3$ thin films tends to increase this value to more than 50 meV \cite{D'Innocenzo:2014,Wu:2014}. 

While the variation of experimental results causes ongoing debate of the excitonic character of the absorption edge, first-principles theoretical spectroscopy can provide deeper understanding.
To this end, Bokdam \emph{et al.}\ used many-body perturbation theory (MBPT) and solved the Bethe-Salpeter equation (BSE) for the optical polarization function, reporting an exciton binding energy of 45 meV in tetragonal MAPbI$_3$ \cite{Kresse_2016}.
Similarly, Zu \emph{et al.}\ computed 40 meV \cite{zhu_computed_2014}
and Umari \emph{et al.}\ computed 30 meV using a similar framework \cite{Umari:2018}.
All three studies attribute the dielectric screening of the electron-hole Coulomb interaction exclusively to electronic interband transitions, corresponding to a high-frequency dielectric constant $\varepsilon_\infty$ of MAPbI$_3$ between 5 and 7\cite{Kresse_2016,Umari:2018,zhu_computed_2014}.
In another work Ahmed \emph{et al.}\ used the BSE framework to predict a binding energy of 153 meV.
While these calculations were done on a coarse 4$\times$4$\times$4 $\bf{k}$-point grid, likely leading to an overestimate of the binding energy \cite{Ahmed:2014}, the values of 45 meV \cite{Kresse_2016} and 40 meV \cite{zhu_computed_2014} quoted above still overestimate experimental data.

However, exciton binding energies are critically influenced by the strength of the electron-hole interaction and, thus, dielectric screening in the material, both in experiment and calculations.
This is important because the lattice structure of MAPbI$_3$ is very polarizable, leading to a large \emph{static} dielectric constant, possibly contributing to screening. 
To this end, Frost \emph{et al.}\ showed that the static dielectric constant of 25.7, accounting for lattice and electronic polarizability, leads to an exciton binding energy of less than 1 meV in a Wannier-Mott model \cite{Frost_2014}.
Evens \emph{et al.}\ used a value of $\varepsilon$=11 to demonstrate that including lattice contributions to screening improves agreement with measured room-temperature absorption spectra \cite{Even:2014}.
Men\'endez-Proupin \emph{et al.}\ use a parabolic-band with a Pollman-B\"uttner type model for polaron screening and find an exciton binding energy of 24 meV \cite{Menéndez-Proupin:2015}.
Umari \emph{et al.}\ also showed that including polar phonon screening reduces the binding energy from 30 meV to 15 meV \cite{Umari:2018}.
Finally, Hakamata \emph{et al.}\ employed non-adiabatic molecular dynamics to calculate the time-averaged exciton binding energy in a dynamical MAPbI$_3$ lattice, predicting a binding energy of 12 meV and a dielectric constant between $10$ and $15$, in excellent agreement with RT measured values \cite{Hakamata:2016}. 
Bokdam \emph{et al.}\ provide arguments against the importance of lattice screening for exciton binding energies of MAPbI$_3$ and instead invoke formation of polarons \cite{Kresse_2016}.

In this work, we study the complementary problem of an additional screening contribution due to free electrons, arising from defects or donors in a sample.
First-principles studies of multiple point defects in MAPbI$_3$ showed that charged defects with low formation energy occupy shallow levels relative to the band extrema \cite{Yin:2014:2,Kim:2014,Liu:2018,Yang:2015:2}.
Wang \emph{et al.}\ showed that synthesis with varying ratios of PbI$_2$:MAI precursors can change samples from $p$- to $n$-type, with free-electron concentrations as high as $3.5\times 10^{18}$ cm$^{-3}$ and even at a standard precursor ratio of 1:1 moles of PbI$_2$ and MAI, a free-electron concentration of $1.8\times 10^{17}$ cm$^{-3}$ was measured \cite{Wang_2014}.
Other studies confirmed free-carrier concentrations in the range of $10^{17}$\,--\,$10^{18}$ cm$^{-3}$ \cite{Dymshits:2015,Guerrero:2014}.
Dielectric screening due to free electrons has been shown to reduce the strength of the electron-hole Coulomb interaction in ZnO \cite{Schleife:2011_d,Schleife:2012,Kang:2019} and, together with Pauli blocking lead to formation of Mahan excitons \cite{Mahan:1967} at the absorption edge.

We speculate that these effects also affect exciton binding MAPbI$_3$ and to clarify this, we perform accurate first-principles simulations of electronic structure and optical properties of MAPbI$_3$.
The remainder of the paper is organized as follows.
Section \ref{sec:methods} summarizes the theoretical and computational approach for solving the BSE to calculate optical response.
Section \ref{sec:results} details results for atomic geometries, electronic structure, and optical properties.
We compute exciton-binding energies and optical spectra, explicitly including  various concentrations of free electrons that arise in the material for varying defect concentrations.
Finally, Sec.\ \ref{sec:conclusions} summarizes and concludes this work.

\section{\label{sec:methods}Computational approach}

We use density functional theory (DFT) \cite{Kohn_Sham_equation,Hohenberg_1964} to compute fully relaxed atomic geometries of the three experimentally most relevant polymorphs of MAPbI$_3$, i.e.\ the orthorhombic (O), tetragonal (T), and cubic (C) phase. 
Their Brillouin zones (BZs) are sampled using $\Gamma$-centered 4\,$\times$\,4\,$\times$\,4, 4\,$\times$\,4\,$\times$\,4, and 6\,$\times$\,6\,$\times$\,6 $\mathbf{k}$-point meshes, respectively.
The projector-augmented wave (PAW) method is used to describe the electron-ion interaction \cite{Blochl:1994} and single-particle wave functions are expanded into a plane-wave basis up to a cutoff energy of 600 eV.
These parameters are sufficient to converge the total energy to within 5 meV per atom.
The PBEsol exchange-correlation (XC) functional \cite{Perdew:2008} has previously been used to predict accurate relaxed atomic geometries for MAPbI$_3$ \cite{Brivio:2015} and is used here for the same purpose.

In order to obtain equilibrium atomic geometries, we initialize the structures of the O, T, and C phases prior to relaxation using those reported in Ref.\  \onlinecite{Brivio:2015}.
This captures the symmetry of ordered MA cations in the O phase and a pseudo-random ordering of the MA sub-lattice in the T phase.
While the C phase exhibits total disordering of the MA cation sub-lattice in experiment \cite{Whitfield:2016}, we study a pseudo-cubic phase with ordered MA cations.
This approach is common in the literature to maintain the uniform alignment of PbI$_3$ octahedra observed experimentally for the C phase\cite{Brivio:2015,Butler:2015,Targhi:2018}.
In experiment the cubic lattice also shows a slight pseudo-cubic behavior, due to rotations of the MA cations \cite{Whitfield:2016}.
We verify that these atomic coordinates correspond to equilibrium structures by computing total energies for several unit-cell volumina within 1\,\% of the equilibrium value and determine the minimum.
All atomic geometries were then relaxed until Hellman-Feynman forces are smaller than 10 meV/\AA.

For these relaxed geometries we compute high- and low-frequency dielectric tensors using density functional perturbation theory (DFPT) \cite{Gajdos:2006} and the generalized-gradient approximation by Perdew, Burke, and Ernzerhof (PBE) \cite{Perdew:1996} to describe XC.
The BZs are sampled using $\Gamma$-centered 5\,$\times$\,5\,$\times$\,5, 5\,$\times$\,5\,$\times$\,5, and 7\,$\times$\,7\,$\times$\,7 $\mathbf{k}$-point meshes for O, T, and C phases, respectively, for these calculations.

In order to compute electronic structures that can be compared to experiment, we  overcome the well-known band gap underestimation of DFT by taking quasiparticle (QP) corrections into account within MBPT \cite{umari_relativistic_2014}.
Furthermore, due to the presence of heavy-metal ions in MAPbI$_3$, spin-orbit coupling (SOC) is included within the PAW approach \cite{Steiner:2016}.
We performed $GW_0$+SOC calculations using 1\,$\times$\,1\,$\times$\,1, 1\,$\times$\,1\,$\times$\,1, and 2\,$\times$\,2\,$\times$\,2 $\Gamma$-centered $\mathbf{k}$-point grids for O, T, and C phases, respectively.
The Green's function was iterated four times to converge QP band gaps to within 25 meV.
4000 empty bands were included for the O and T structures, and 2000 for the C phase.

Finally, we study optical response including excitonic effects by solving the Bethe-Salpeter equation (BSE) for the optical polarization function.
The BSE in the Bloch basis can be written as an eigenvalue equation \cite{Rohfling_BSE,Fuchs_2008} for the Hamiltonian
\begin{equation}
H(cv\mathbf{k},c'v'\mathbf{k'})=\left(\epsilon_{c\mathbf{k}}-\epsilon_{v\mathbf{k}}\right)\delta_{cc{'}}\delta_{vv{'}}\delta_{\mathbf{kk'}}+2v_{c'v'\mathbf{k}'}^{cv\mathbf{k}} - W_{c'v'\mathbf{k}'}^{cv\mathbf{k}}
\end{equation}
The indices $c$, $v$, and $\mathbf{k}$, refer to conduction band, valence band, and point in reciprocal space, respectively.
The term in parentheses on the right-hand side represents single-QP excitation energies of non-interacting electron-hole pairs, described by the QP band structure.
For the QP energies $\epsilon_{c\mathbf{k}}$ and $\epsilon_{v\mathbf{k}}$ we use results computed using PBE+SOC, with the band gap rigidly shifted to the $GW_0$+SOC gap.
Bloch integrals, that enter the exchange interaction
$2v_{c{'}v{'}\mathbf{k}{'}}^{cv\mathbf{k}}$ and the screened Coulomb interaction $W_{c{'}v{'}\mathbf{k}{'}}^{cv\mathbf{k}}$ between electrons and holes, are evaluated using spin-polarized DFT-PBE Kohn-Sham eigenstates.
Electronic interband screening of the electron-hole Coulomb interaction is computed using the model dielectric function proposed by Bechstedt \emph{et al.}\ in the absence of free carriers \cite{Bechstedt:1992,Model_Dielectric_Function} and approximated as a dielectric constant $\varepsilon_\infty$ when free carriers are taken into account \cite{Schleife:2011_d,Schleife:2012}.
Optical spectra of C MAPbI$_3$ are computed over a wide energy range using a 11\,$\times$\,11\,$\times$\,11 $\mathbf{k}$-point grid with a small random shift and for a careful examination of the spectral onset up to an energy of 2.2 eV, a hybrid 5:2:32.5 $\mathbf{k}$-point grid is used (see Ref.\ \onlinecite{Fuchs:2008_b} for nomenclature).
Independent-particle spectra for the O and T phases were computed using 7\,$\times$\,7\,$\times$\,7 $\mathbf{k}$ points with a small random shift.

In order to describe the influence of free electrons, we account for Burstein-Moss shift (BMS), band-gap renormalization (BGR), and additional free-electron screening of the electron-hole interaction \cite{Schleife:2011_d,Schleife:2011_b,Schleife:2010_b,Kang:2019}.
This approach has been successfully used to describe optical properties of doped ZnO before \cite{Schleife:2011_d,Schleife:2012}.
BMS arises from Pauli blocking of the lowest conduction-band states that are occupied by the additional free electrons, and is taken into account via their occupation numbers.
BGR is described as an additional scissor shift, computed using the analytical model of Berggren and Sernelius \cite{Berggren:1981}. 
Finally, the intraband contribution to dielectric screening is described as a free-electron response, which in the limit of small wave vectors becomes 
\begin{equation}
\label{eqn:model_eps_fc}
\varepsilon_\mathrm{fc}(q,n_c) =
\varepsilon_\infty \left( 1+\frac{q^2_\mathrm{TF}}{q^2} \right),
\end{equation}
with the Thomas-Fermi wave vector
\begin{equation}
\label{eqn:eps_eff_qn}
q_\mathrm{TF} =  \sqrt{\frac{3n_c e^2}{2 \varepsilon_0\varepsilon_\infty \epsilon_\mathrm{F}}}.
\end{equation}
Here, $n_c$ is the concentration of free electrons in the system, $\varepsilon_0$ is the vacuum permittivity, $\varepsilon_\infty$ is the directionally averaged high-frequency dielectric constant of the material, and $\epsilon_\mathrm{F}$ is the Fermi level corresponding to $n_c$.

All DFT and $GW$ calculations are carried out using the Vienna \emph{Ab-Initio} Simulation Package \cite{Gajdos:2006,Kresse:1999,Kresse:1996,Shishkin:2006} (VASP).
The BSE calculations are performed using the implementation described in Refs.\ \onlinecite{Roedl:2008,Fuchs:2008_b}.
All input and output of this work is available in the Materials Data Facility \cite{mdf_perosvkite}.

\section{\label{sec:results}Results and Discussion}

\subsection{Atomic Geometries}

\begin{figure}
\includegraphics[width=0.98\columnwidth]{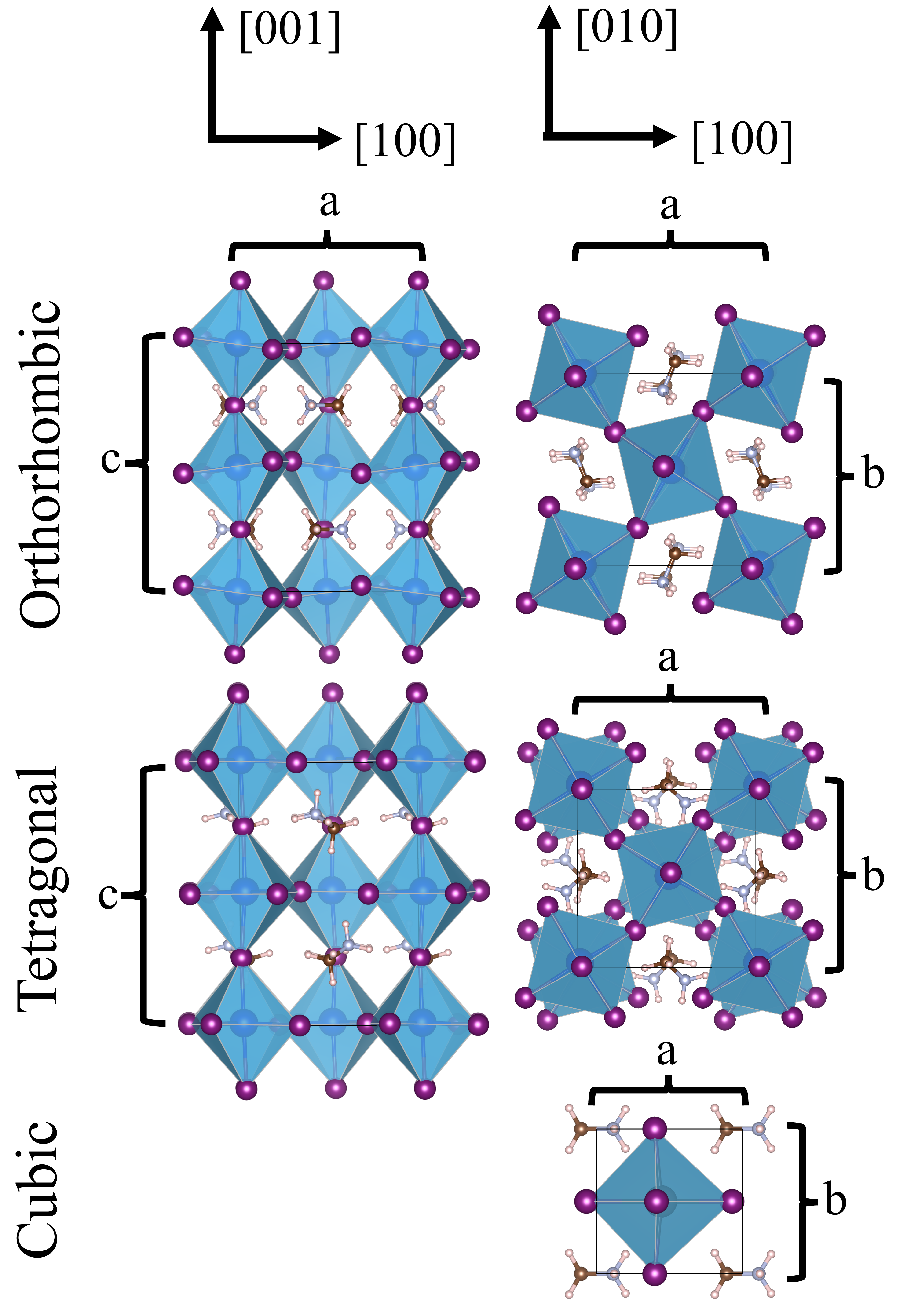}
\caption{\label{fig:strucs}(Color online.)
Relaxed atomic geometries of orthorhombic, tetragonal, and cubic phases of MAPbI$_3$.
Ions are represented as gray (Pb), purple (I), brown (C), pink (H), and blue (N) spheres.
Lattice constants $a$, $b$, and $c$ align with the [001], [010], and [001] directions, respectively.
}
\end{figure}

First, we study relaxed atomic geometries of the low-temperature equilibrium phase of MAPbI$_3$, the orthorhombic (O) crystal structure with space group Pnma \cite{ong_phases_2015,Whitfield:2016}.
This phase has minimum entropy by ordering CH$_3$NH$_3$ ions periodically \cite{Brivio:2015} and the PbI$_3$ sub-lattice forms stacked octahedra that are tilted with respect to the [001] axis of the unit cell (see Fig.\ \ref{fig:strucs}).
Angles between lattice vectors are all 90$^{\circ}$ and the lattice parameters are non-equal, with $a$=8.37, $b$=9.07, and $c$=12.67 \AA. 
The $c$ axis agrees well with experimental values between 12.1\,--\,12.6 \AA\ and the relaxed $a:b$ aspect ratio in this work of 0.921 only slightly underestimates that seen in experiments, 0.97\,--\,0.98\cite{EXP_PARAM_PbI3,Whitfield:2016}.

Experiment also shows that above $T$=162 K, MAPbI$_3$ undergoes a phase transformation to a tetragonal (T) phase with space group I4/mcm (see Fig.\ \ref{fig:strucs}) \cite{ong_phases_2015,Whitfield:2016}.
This first-order phase transition is marked by three phenomena:
First, we compute a change in lattice parameters from $a\neq b \neq c$ in the O to $a$=8.70, $b$=8.72, and $c$=12.83 \AA\ in the T phase.
The relaxed structure results in good agreement between $a$ and $b$ with a difference of only $\approx 0.02$ \AA.
Second, there is disordering of CH$_3$NH$_3$ ions in the T phase, that leads to a disordered cation sub-lattice.
To approximate this effect in our unit cell, we disorder the organic cations, based on the structures of Brivio \emph{et al.}\ \cite{Brivio:2015}
This disorder is stabilized by $c$-axis tilting in the T phase.
Finally, alternating tilts of the octahedrons in the [001] direction appear, which in turn stabilizes the $a$=$b$ condition  \cite{ong_phases_2015,Whitfield:2016}.

At even higher temperatures above $T$=327 K, T MAPbI$_3$ undergoes another transition to a cubic (C) phase with space group Pm$\bar{3}$m (see Fig.\ \ref{fig:strucs}) \cite{ong_phases_2015}.
This C phase is stabilized through total disordering of the MA cation sublattice.
Since thermal rotation of MA cations is not accounted for in the geometry relaxation \cite{Whitfield:2016}, we follow the common approach of modelling this phase as a pseudo-cubic distortion of the Pm$\bar{3}$m cubic perovskite structure with ordered MA cations \cite{ong_phases_2015,Whitfield:2016}.
This lattice geometry is slightly triclinic;
in experiment, it is also pseudo-cubic due to rotations of the MA cations \cite{Whitfield:2016}.
Relaxed atomic geometries result in slightly tilted axes compared to the experimental Pm$\bar{3}$m phase, which agrees with earlier computational reports:
Ong \emph{et al.}\ showed that in DFT calculations the distorted C phase (space group $P4mm$) is more stable compared to a constrained Pm$\bar{3}$m phase \cite{ong_phases_2015}.
The average of the pseudo-cubic lattice constants $(a+b+c)/3$=6.31 \AA\ agrees well with measurements \cite{Whitfield:2016}.

Overall, our results for relaxed atomic geometries are in excellent agreement with values from experiment and previous calculations.
A more detailed comparison to other work can be found in Table S1 of the supplemental material.

\subsection{\label{subsec:electronic}Electronic Structure}

\begin{table}
\caption{\label{tab:gaps}
Band gaps $E_\mathrm{g}$ (in eV), static ($\varepsilon_0$) and static electronic ($\varepsilon_\infty$) dielectric constants, and effective electron ($m_c$) masses for MAPbI$_3$ from our calculations and the literature.
Results from $G_0W_0$, $GW_0$ (iteration only of the Green's function), SS-$GW$ (self-consistent scissor $GW$\cite{Filip:2014}), and QS$GW$ (quasiparticle self-consistent $GW$) are shown for $E_\mathrm{g}$. Density-functional perturbation theory (DFPT) and molecular dynamics (MD) are compared for dielectric constants.
}
\begin{tabular}{ c c c c }
\hline
 & Orthorhombic & Tetragonal & Cubic \\
\hline
$E_\mathrm{g}$ (PBE) & 1.55 & 1.40 & 1.51 \\
$E_\mathrm{g}$ (PBE+SOC) & 0.66 & 0.70 & 0.56 \\
$E_\mathrm{g}$ ($GW_0$+SOC) & 1.42 & 1.39 & 1.38 \\ 
\hline
$E_\mathrm{g}$ (PBE) & 1.61\cite{Thind:2017} & 1.45\cite{Thind:2017}, 1.68 \cite{umari_relativistic_2014}  & 1.44\cite{Thind:2017} \\
$E_\mathrm{g}$ ($G_0W_0$+SOC) & 1.81 \cite{Quarti:2016}, 1.32\cite{Filip:2014} & 1.62 \cite{umari_relativistic_2014}, 1.67\cite{Quarti:2016} & 1.28 \cite{Quarti:2016}, 1.48\cite{Ahmed:2014} \\
$E_\mathrm{g}$ (SS-$GW$+SOC) & 1.79\cite{Filip:2014,Filip:2015} & -- & -- \\
$E_\mathrm{g}$ (QS$GW$+SOC) & -- & -- & 1.67\cite{Brivio:2014} \\
$E_\mathrm{g}$ (Exp.) & 1.65\cite{Quarti:2016} & 1.5\,--\,1.61 \cite{Quarti:2016} & 1.69 \cite{Quarti:2016} \\
\hline
\hline
$\varepsilon_{\infty}$ (DFPT) & 6.22 & 6.23 & 6.24 \\
$\varepsilon_{0}$ (DFPT) & 23.17 & 22.66 & 22.1\\
\hline
$\varepsilon_{\infty}$ (DFPT) & 5.80 \cite{PerezOsorio_2015} & 5.50 \cite{umari_relativistic_2014}, 6.60\cite{Umari:2018} & 6.83 \cite{Kresse_2016} \\ 
$\varepsilon_{\infty}$ (Exp.) & -- & 5.00\cite{Sendner:2016} & -- \\ 
$\varepsilon_{0}$ (DFPT) & 25.30\cite{PerezOsorio_2015} & -- & --\\
$\varepsilon_{0}$ (MD) & -- & 30.00\cite{Kresse_2016} & --\\
$\varepsilon_{0}$ (Exp.) & -- & 33.50\cite{Sendner:2016}, 28.80\cite{Poglitsch:1987} & --\\
\hline
\hline
$m_{c}$ (PBE+SOC) & 0.19 & 0.16 & 0.23\\
\hline
$m_{c}$ (DFT+SOC) & 0.19\cite{Feng:2014}, 0.11\cite{Filip:2015} &   0.15\cite{Frost_2014} & 0.23 \cite{Giorgi:2013} \\
$m_{c}$ ($G_0W_0$+SOC) & 0.16\cite{Filip:2015}  & 0.17 \cite{umari_relativistic_2014} & -- \\
$m_{c}$ (SS-$GW$+SOC) & 0.21\cite{Filip:2015}  & --  & -- \\
\hline
\end{tabular}
\end{table}

\begin{figure}
\includegraphics[width=0.98\columnwidth]{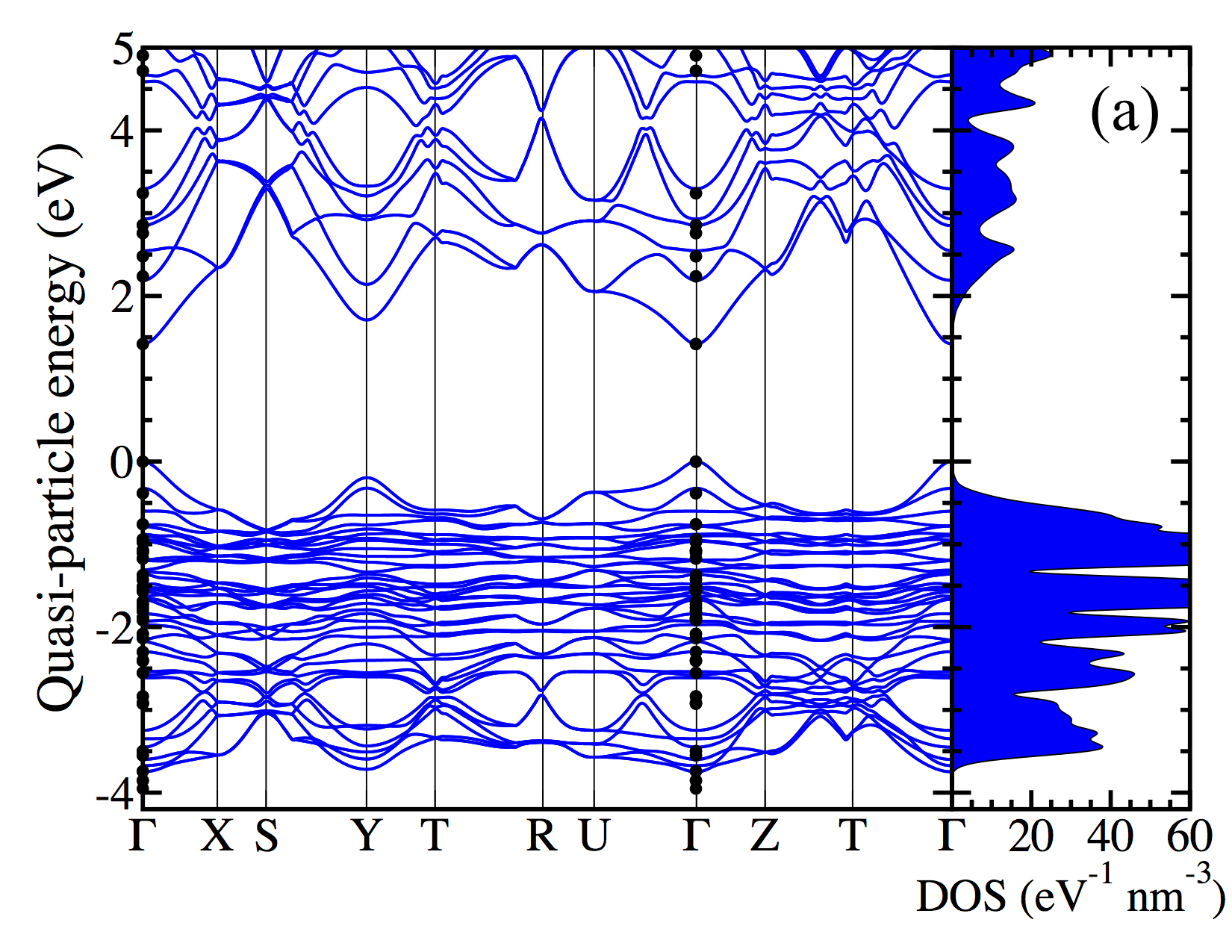}
\includegraphics[width=0.98\columnwidth]{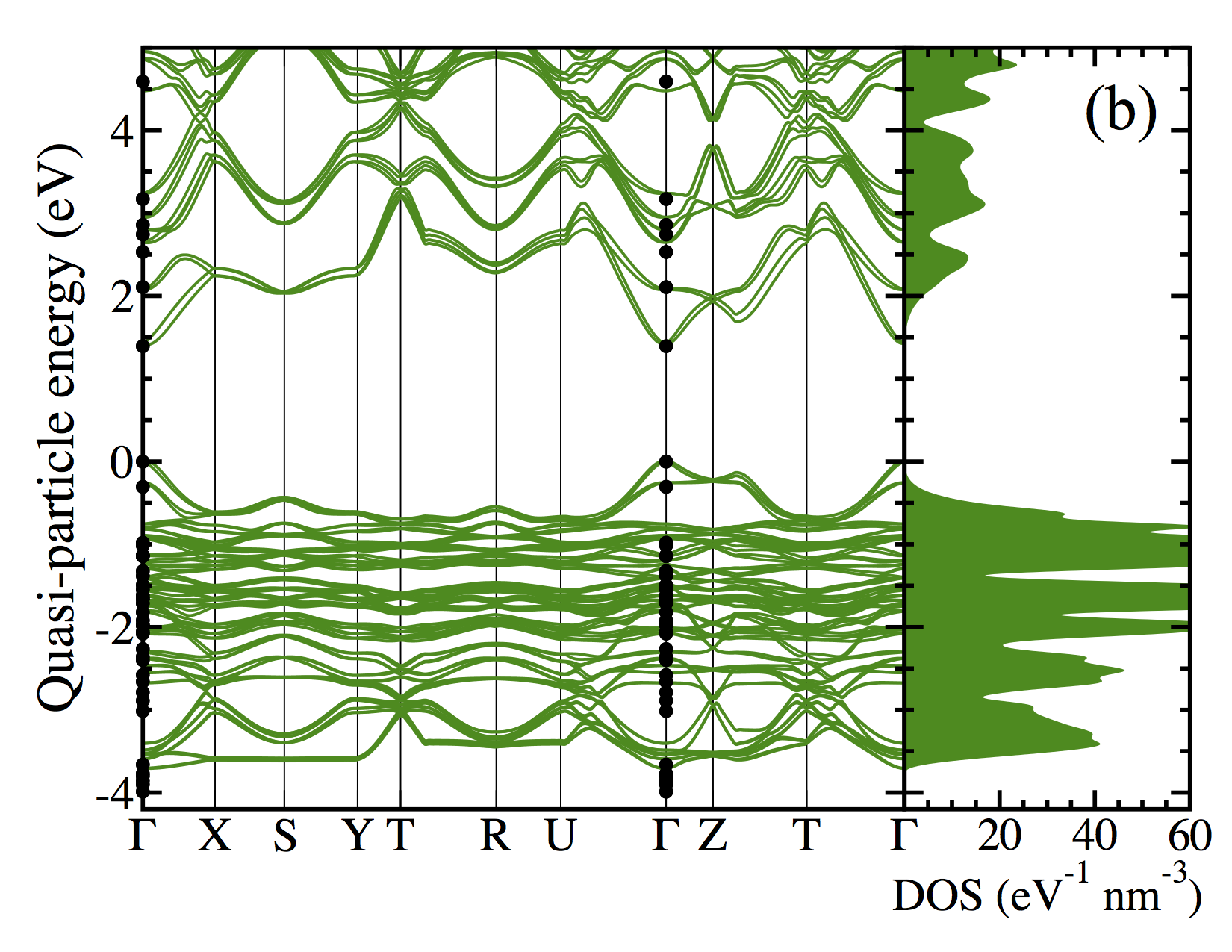}
\includegraphics[width=0.98\columnwidth]{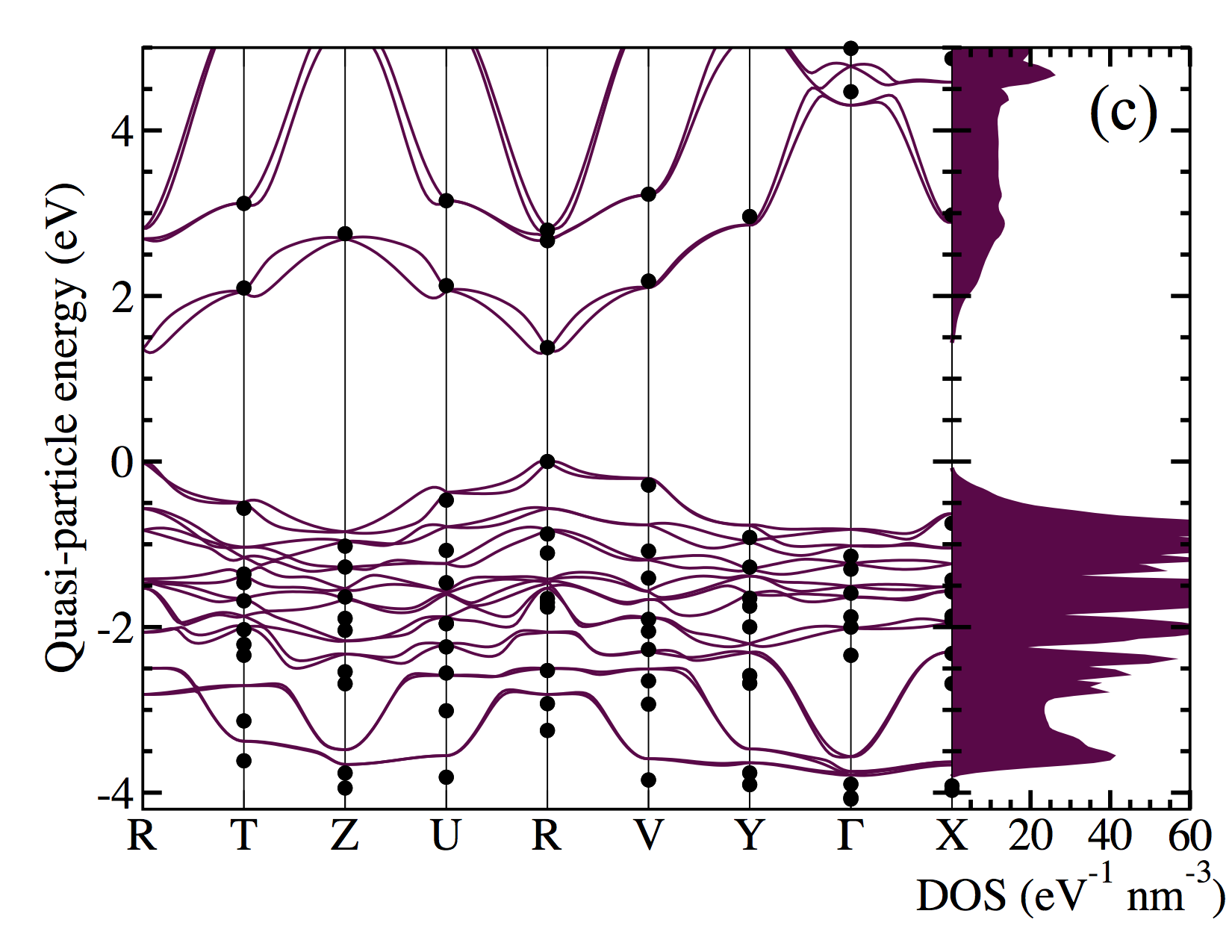}
\caption{\label{fig:bands}(Color online.)
Kohn-Sham band structure and density of states from PBE+$\Delta_{GW_0}$+SOC (solid lines) and $GW_0$+SOC (black circles) calculations for orthorhombic (a), tetragonal (b), and cubic (c) phases of MAPbI$_3$.
All conduction states are rigidly shifted to the $GW_0$+SOC gap. The valence-band maximum is used as energy zero.
}
\end{figure}

Using the $GW_0$+SOC approach, we compute band gaps of 1.42, 1.39, and 1.38 eV, for the O, T, and C phase, respectively (see Table \ref{tab:gaps}).
Figure \ref{fig:bands} shows direct band gaps for each phase that are located at the $\Gamma$ point of the BZ for O and T phase, and at the $R$ point for the C phase.
This change in reciprocal-space location of the direct gap is a consequence of cell symmetry \cite{Pedesseau:2016}.
Our results for MAPbI$_3$ band gaps are consistent with previous $GW$ calculations and only slightly underestimate experimental values of 1.5\,--\,1.7 eV (see Table \ref{tab:gaps}).
This table also shows that previous calculations produced varying results based on the specific $GW$ approximation and description of SOC \cite{even_importance_2013,umari_relativistic_2014,Mosconi:2016,Perovskite_Review_2015}.
In particular, Filip and Giustino showed that different schemes for including relativistic effects and iterating the $GW$ method resulted in different values for the gap:
Using fully relativistic pseudopotentials for Pb and I and the scissor-self consistent $GW$ method \cite{Filip:2014} to iterate QP wave functions and eigenenergies, they predicted 1.79 eV for the orthorhombic phase \cite{Filip:2014}.
Separately, Umari \emph{et al.}\ reported 1.62 eV for the T phase \cite{umari_relativistic_2014}. 

As expected, gaps at the PBE+SOC level of theory severely underestimate experimental results by more than $1$ eV for each phase.
Using our $GW_0$+SOC data, we can correct this for the calculation of optical spectra, using a rigid scissor shift;
we denote this approach by PBE+$\Delta_{GW_0}$+SOC.
Figure \ref{fig:bands} compares band structures of O, T, and C MAPbI$_3$ at the PBE+$\Delta_{GW_0}$+SOC and $GW_0$+SOC levels of theory and illustrates the density of states (PBE+$\Delta_{GW_0}$+SOC).
As can be seen in Fig.\ \ref{fig:bands} the direct nature of the gap is broken by a strong Rashba-Dresselhaus spin-orbit splitting for C MAPbI$_3$.
The effect is smaller for T MAPbI$_3$ and has been studied extensively for T and C phases \cite{Mosconi:2016,Etienne:2016,Mosconi:2017}.

By comparing $GW_0$+SOC energies at high-symmetry $\mathbf{k}$ points to the electronic structure from PBE+$\Delta_{GW_0}$+SOC in Fig.\ \ref{fig:bands}, we illustrate for C MAPbI$_3$ that the latter is a suitable basis for optical calculations.
Here we are interested in optical response in the visible spectral range, hence, we focus on electronic states within 1.6 eV of the band extrema.
As can be seen in Fig.\ \ref{fig:bands}(c) the conduction band dispersions from both approaches are in excellent agreement in this energy range:
While the direct gap appears at $R$, the largest deviation for the lowest conduction band amounts to about 0.15 eV and occurs at the $\Gamma$-point.
Overall, the valence bands are also in good agreement between both approaches. PBE+$\Delta_{GW_0}$+SOC results
tend to predict valence band energies only slightly higher in energy than those predicted by $GW_0$+SOC, for instance 0.4 eV at the $\Gamma$ point [see Fig.\ \ref{fig:bands}(c)].
The overall width of the uppermost valence block is 0.35 eV larger at the $GW_0$+SOC level of theory.
Hence, overall, our data indicates that excitation energies are underestimated by at most 0.3\,--\,0.4 eV when computing optical spectra starting from the PBE+$\Delta_{GW_0}$+SOC electronic structure.
Finally, effective electron masses are determined by a parabolic fit near the band edge of our PBE+$\Delta_{GW_0}$+SOC data and also reported in Table \ref{tab:gaps}.

\subsection{\label{optics_no_fc}Optical Response: Independent-quasiparticle approximation}

\begin{figure}
\includegraphics[width=1.0\columnwidth]{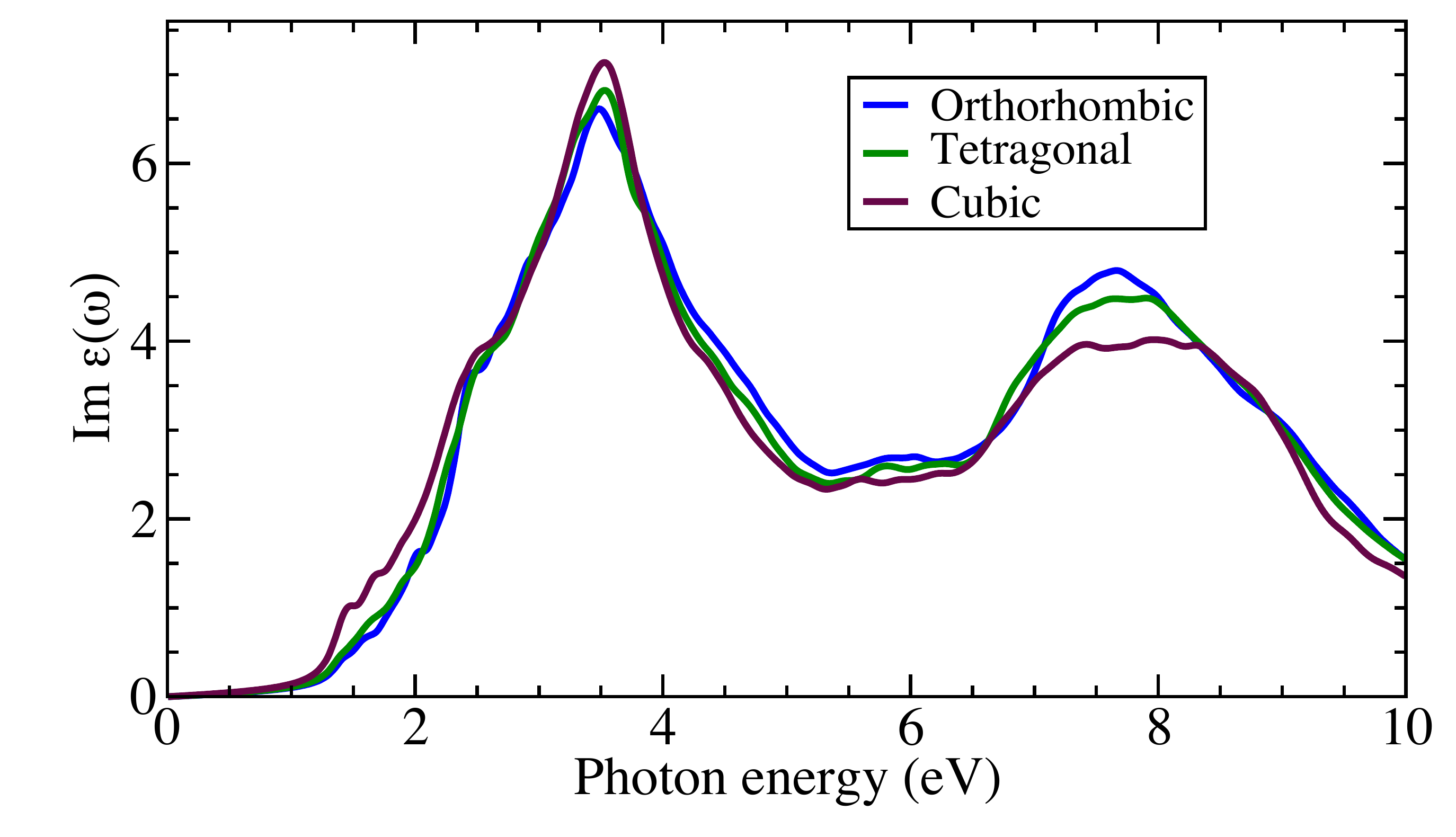}
\caption{\label{fig:SPECS_PHASES}(Color online.)
Polarization-averaged imaginary part of the frequency-dependent dielectric functions of MAPbI$_3$, calculated within independent-quasiparticle approximation using PBE+$\Delta_{GW_0}$+SOC.
Blue, green, and maroon curves correspond to orthorhombic, tetragonal, and cubic phase, respectively.
}
\end{figure}

The optical absorption spectra of all three MAPbI$_3$ phases share similar spectral features, as shown in Fig.\ \ref{fig:SPECS_PHASES}.
In this figure, we illustrate the polarization-averaged imaginary parts of the dielectric functions, computed using the independent-quasiparticle approximation within PBE+$\Delta_{GW_0}$+SOC.
Our results agree overall well in the visible region between 1.5 and 3 eV with the absorption coefficient calculated using fully-relativistic $G_0W_0$+SOC \cite{umari_relativistic_2014} for T MAPbI$_3$, as shown explicitly in Fig. S4 of the supplemental material.

Figure \ref{fig:SPECS_PHASES} shows a smooth, gradual onset of absorption at the $GW_0$ band gap for all three phases. At higher energies near 2.4 eV all spectra show a shoulder feature which we attribute to optical transitions between the uppermost valence band and lowest conduction band at $\mathbf{k}$ points slightly away from the location of the band extrema (see Fig. S2 in the supplemental material).
The difference of the lowest conduction and highest valence bands shows that transitions near the $\Gamma$, $Y$, and $U$ point (orthorhombic), near the $\Gamma$ and $S$ point (tetragonal), and near the $T$, $U$, and $V$ point (cubic) dominate between 2.2 and 2.6 eV. 
From the shoulder, $\varepsilon_2$ further increases into the UV energy region and peaks at 3.48, 3.53, and 3.53 eV for O, T, and C phase, respectively.
The major contributions to this peak are optical transitions between the highest valence band and lowest conduction band at $\mathbf{k}$ points far from the location of the band extrema, e.g.\ the $Z$ point in C MAPbI$_3$.
Fig. S3 in the supplemental information also indicates that there are minor contributions from transitions from lower valence bands into the lowest conduction band.
Our assignment of these spectral features agrees with that in Ref.\ \onlinecite{Herz:2016}. 
Finally, after this peak optical response, $\varepsilon_2$ decreases until about 5.3 eV and then increases again gradually to a much broader peak, centered around 7.7 eV, which is far outside the visible spectrum.

We also computed the static ($\varepsilon_0$) and static electronic ($\varepsilon_\infty$) dielectric constants of MAPbI$_3$ using DFPT and the PBE electronic structure.
For $\varepsilon_\infty$ we find very similar values around 6.23 for all three phases (see Table \ref{tab:gaps}).
Our results are in the midst of previously calculated and measured values ranging from 5.5 to 7.0 \cite{PerezOsorio_2015,umari_relativistic_2014,Kresse_2016,Manser:2014}.
We confirmed that the same magnitude but opposite sign of quasi-particle and SOC induced shifts, reported before for band gaps \cite{umari_relativistic_2014}, justifies using DFPT based on the PBE electronic structure to compute dielectric constants.
Due to the large lattice polarizability of MAPbI$_3$, our DFPT results for $\varepsilon_0$ are much larger than $\varepsilon_{\infty}$, with values of 22.1\,--\,23.2 for the three phases (see Table \ref{tab:gaps}).
These results are in good agreement with earlier data from DFPT and molecular dynamics simulations, as well as experimental measurements, in the range of 25\,--\,35\cite{Sendner:2016,Poglitsch:1987}.

\subsection{\label{optics_exc}Optical Response: Excitonic effects}

\begin{figure}
\includegraphics[width=1.0\columnwidth]{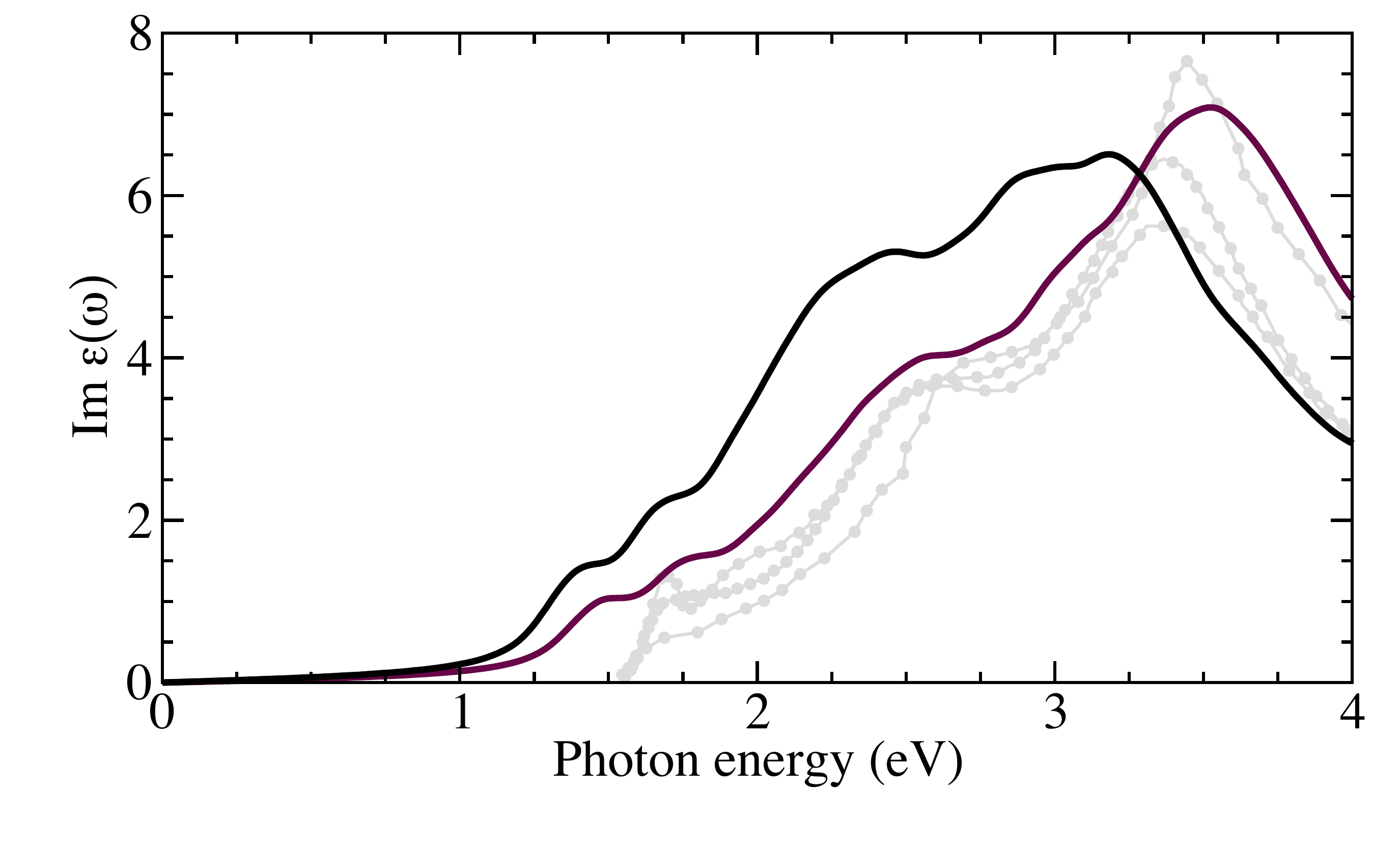}
\caption{\label{fig:SPECS_noeocc}(Color online.)
Polarization-averaged imaginary part of the frequency-dependent dielectric function of cubic MAPbI$_3$.
Results from independent-quasiparticle approximation, here PBE+$\Delta_{GW_0}$+SOC (solid maroon line), are compared to the BSE$_\mathrm{el}$+$\Delta_{GW_0}$+SOC approach (solid black line) that accounts for electron-hole interaction, and to experimental results (gray diamonds) \cite{exp_spec,Chen_CW_2015,Loper_2014}.
}
\end{figure}

Next, we study the influence of excitonic effects on optical absorption of MAPbI$_3$.
To this end, Fig.\ \ref{fig:SPECS_noeocc} compares the independent-quasiparticle spectrum to the solution of the BSE, accounting for electronic interband screening as described by a model dielectric function \cite{Bechstedt:1992,Model_Dielectric_Function} parameterized using a dielectric constant of $\varepsilon_\infty$=6.24 (see Table \ref{tab:gaps}).
Given the similarities of the independent-quasiparticle optical spectra of the three different phases (see Fig.\ \ref{fig:SPECS_PHASES}), we only focus on C MAPbI$_3$ in the following.

For C MAPbI$_3$, Fig.\ \ref{fig:SPECS_noeocc} shows that excitonic effects cause a significant red shift of the absorption onset and of higher-energy features.
While the onset of the independent-quasiparticle spectrum occurs at the $GW_0$+SOC band gap of 1.38 eV (see Table \ref{tab:gaps}), the lowest eigenvalue of the BSE Hamiltonian is about 64.5 meV lower in energy.
Note, that this value is not a well-converged result for the exciton binding energy due to \textbf{k}-point sampling, as we discuss below \cite{Fuchs_2008}.
Energy positions of higher-energy spectral features show larger excitonic shifts;
for instance, the main peak around 3.5 eV in independent-quasiparticle approximation red-shifts by about 0.5 eV.
The shift is accompanied by a redistribution of spectral weight:
When including excitonic effects, features at lower energies are amplified, thus increasing the amplitude of optical absorption at lower energies.

Interestingly, in Fig.\ \ref{fig:SPECS_noeocc} positions of peaks and shoulders in the experimental spectrum seem to agree better with the independent-quasiparticle spectrum.
However, we emphasize the notable difference of about 0.2\,--\,0.3 eV of the absorption onsets that is apparent in the figure and originates from the slightly smaller $GW_0$+SOC gap, compared to experiment.
If this is corrected for, e.g.\ by rigidly shifting the absorption onset to the experimental value, we find excellent agreement of the BSE$_\mathrm{el}$+$\Delta_{GW_0}$+SOC result with experiment across the entire energy range, while the independent-particle spectrum then overestimates the position of the main peak around 3.5 eV by about 0.2\,--\,0.3 eV.
In the following, we analyze how the description of the optical response changes in the presence of free electrons.

\subsection{Optical Response: Free electrons}

\begin{figure}
\includegraphics[width=1.0\columnwidth]{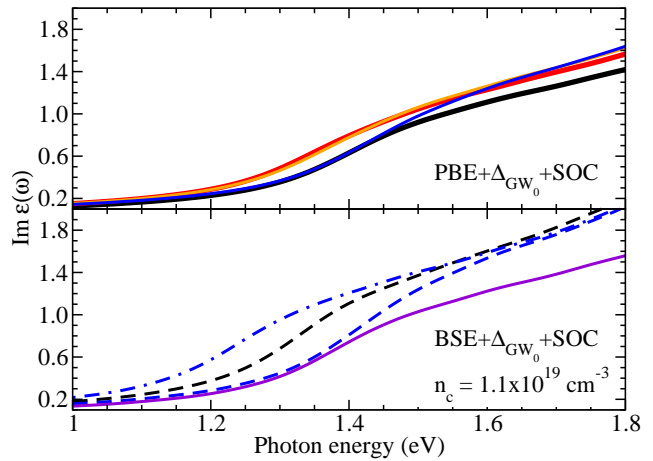}
\caption{\label{fig:BM_shift_specs}(Color online.)
Polarization-averaged imaginary part of the frequency-dependent dielectric function of C MAPbI$_3$ without free electrons (black), and with free-electron concentrations of $2.3\times 10^{18}$ cm$^{-3}$ (red), $5.0\times 10^{18}$ cm$^{-3}$ (orange), and $1.1\times 10^{19}$ cm$^{-3}$ (blue).
A dense, hybrid 5:2:32.5 $\mathbf{k}$-point grid was used.
The top panel shows the influence of BMS and BGR on the independent-quasiparticle spectrum (PBE+$\Delta_{GW_0}$+SOC).
The bottom panel demonstrates the influence of free electrons, $n_c$=$1.1\times 10^{19}$ cm$^{-3}$, on excitonic effects.
The BSE$_\mathrm{el}$+$\Delta_{GW_0}$+SOC spectrum without free electrons (black dashed line) is compared to data that includes free-electron screening without (blue dot-dashed line) and with (blue dashed line) Pauli blocking.
The violet curve approximately decribes lattice screening via the low-frequency dielectric constant $\varepsilon_0$=22.1 in the model dielectric function \cite{Model_Dielectric_Function,bechstedt_ions}.
}
\end{figure}

\begin{figure}
\includegraphics[width=1.0\columnwidth]{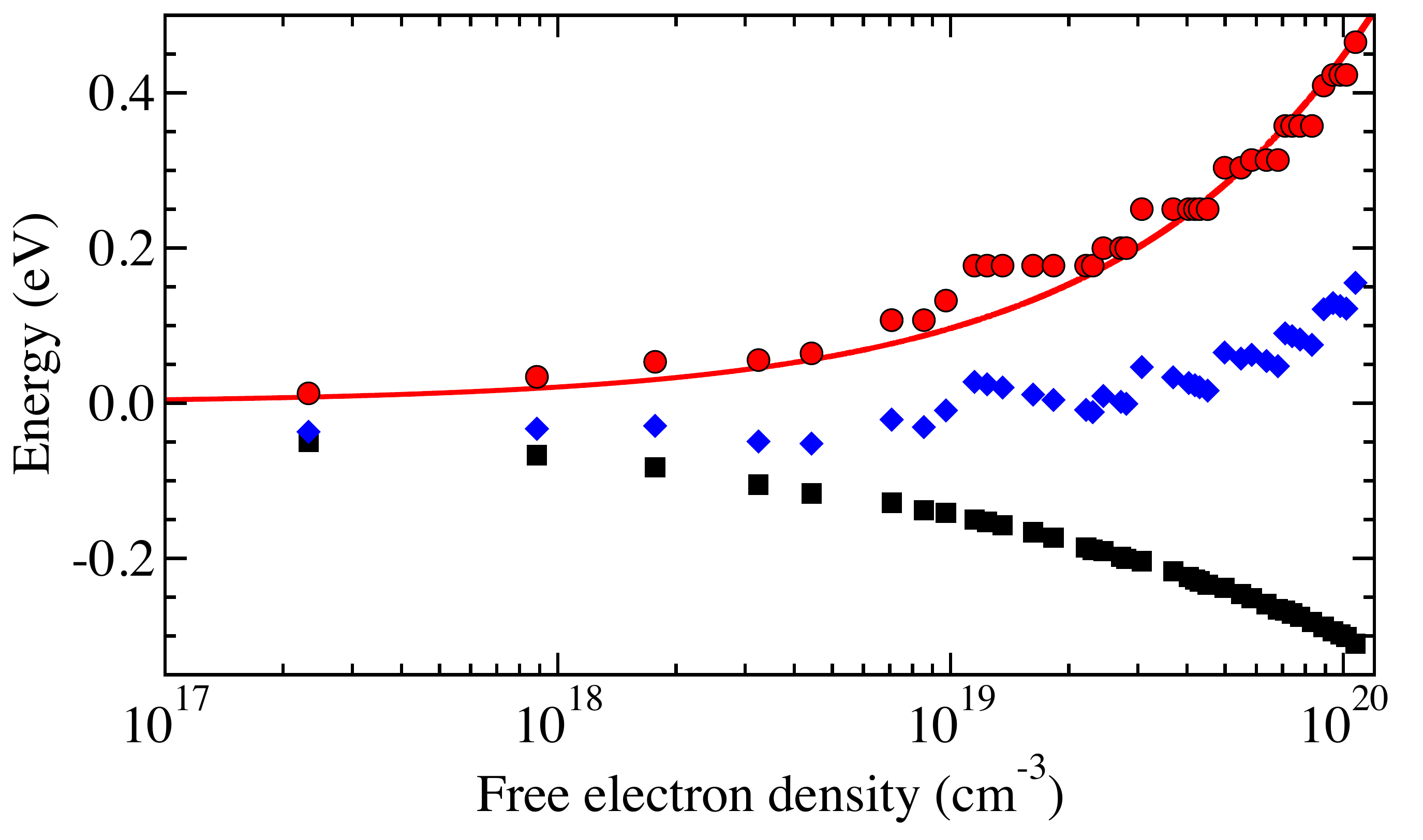}
\caption{\label{fig:BM}(Color online.)
Burstein-Moss shift (red circles), band gap renormalization (black squares), and sum of both (blue diamonds) as a function of free-electron concentration in the conduction band of C MAPbI$_3$.
A dense, hybrid 5:2:32.5 $\mathbf{k}$-point grid was used.
The red line is a curve fit to the BMS data of the form $E_\mathrm{BMS}$=$An_c^{3/2}$.
}
\end{figure}

We first study how Burstein-Moss shift (BMS) and band-gap renormalization (BGR), i.e.\ two effects attributed to free electrons in the conduction band of C MAPbI$_3$, affect the independent-quasiparticle optical spectrum (see top panel of Fig.\ \ref{fig:BM_shift_specs}).
The predicted BMS due to Pauli blocking of optical transitions for a free-electron density of $10^{17}$ cm$^{-3}$ is less than 2 meV
and only reaches a value of about 10 meV for $10^{18}$ cm$^{-3}$  (see Fig.\ \ref{fig:BM}).
Realistic intrinsic $n$ or $p$ type shallow defect concentrations or free-electron-hole densities under illumination \cite{Manser:2014} fall within the range of $10^{15}$ cm$^{-3}$ to $10^{17}$ cm$^{-3}$ and we conclude that for these BMS is only a minor factor.
However, we note that high-intensity illumination has produced free-carrier concentrations around $10^{19}$ cm$^{-3}$\cite{Manser:2014}.
In Fig.\ \ref{fig:BM} we show that in this regime BMS can be on the order of 0.1 eV and quickly increases thereafter, approximately following a $n_c^{3/2}$ dependence.

At the same time, Fig.\ \ref{fig:BM} also illustrates that BGR is on the same order of BMS for C MAPbI$_3$, but with an opposite sign.
As a result, these two effects compensate each other to very high accuracy across an unusually large free-electron range, up to about $10^{19}$ cm$^{-3}$.
This explains why experimental observation of BMS+BGR at the absorption edge \cite{Manser:2014,Fu:2017} requires very high free-carrier concentrations:
Valverde \emph{et al.}\ do not explicitly report \cite{Valverde:2015} any effect of BMS or BGR at a free-carrier density of about $3.3\times 10^{17}$ cm$^{-3}$.
Manser \emph{et al.}\ report\cite{Manser:2014} a rise of the onset by about 0.08 eV for $n_c$=$1.5\times 10^{19}$ cm$^{-3}$, which is between our result for BMS and BMS+BGR.

\begin{figure}
\includegraphics[width=0.98\columnwidth]{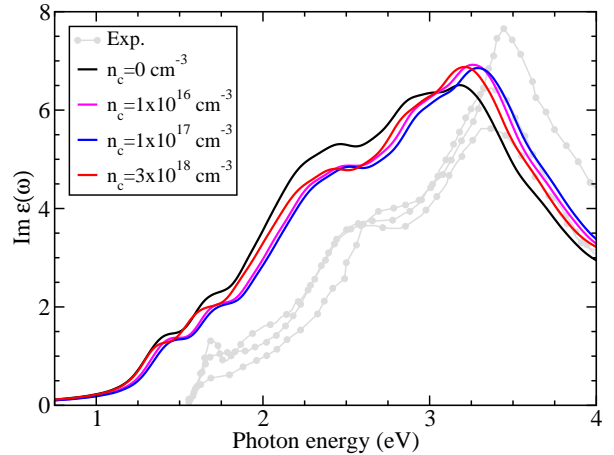}
\caption{\label{fig:exp_comparison}(Color online.)
Polarization-averaged imaginary part of the frequency-dependent dielectric function, computed using the BSE$_\mathrm{el+fc}$+$\Delta_{GW_0}$+SOC approach to account for excitonic effects.
Results are shown for three different experimentally relevant free-electron concentrations and compared to data without free electrons and experiment \cite{exp_spec,Chen_CW_2015,Loper_2014}.
As BMS and BGR are negligible for small values of $n_c$, they are only included for $n_c$=$3\times10^{18}$ cm$^{-3}$.
}
\end{figure}

Next, in order to describe the influence of free electrons on the electron-hole interaction and, thus, excitonic effects, we describe electronic interband screening by a dielectric constant and include free-electron screening using Eq.\ \eqref{eqn:model_eps_fc} when solving the BSE.
Figure \ref{fig:exp_comparison} compares the resulting imaginary part of the dielectric function of C MAPbI$_3$ without free electrons to results for three different free-electron concentrations.
While we find a blue shift of about 0.2 eV for the main absorption peak around 3 eV, interestingly the absorption onset is almost unaffected by free electrons, both in terms of energy position and line shape.
The energy position of the absorption onset barely changes since (i) BMS and BGR largely compensate each other over a large range of free-electron concentrations (see Fig.\ \ref{fig:BM}), and (ii) at the same time, the exciton binding energy is small already in the system without free-electrons.
Hence, its reduction in the presence of free electrons and the formation of a Mahan exciton, does not lead to significant shifts of the absorption edge.
Below we discuss that this Mahan exciton is also the reason why the absorption line shape barely changes in the system with free electrons.

In addition, in Fig.\ \ref{fig:exp_comparison} we compare to three experimental results \cite{Loper_2014,Chen_CW_2015,exp_spec}.
These show good consistency for the major spectral features, i.e.\ the onset at 1.55 eV, the shoulder at 2.62\,--\,2.69 eV, and the peak at 3.35\,--\,3.44 eV.
These peaks and shoulders are reproduced well in our simulations.
The only major difference is that the computed spectra appear red shifted with respect to experiment, which above we attributed to the difference of the single-QP band gaps (see Table \ref{tab:gaps}).
The optical absorption band width, captured by the energy difference of absorption onset and main peak, is 0.15 eV larger when free electrons are present and in slightly better agreement with experiment than the spectrum without free electrons.
Also, the ratio of the dielectric function at the main peak and the shoulder at about 0.5 eV lower energies of 0.53, 0.58, and 0.66 in experiment \cite{Loper_2014,Chen_CW_2015,exp_spec}, improves from 0.82 without free electrons to about 0.7 when accounting for free electrons.
Another notable feature is the narrowing of the spectral peak width when free electrons are included, improving agreement with experiment. 

Finally, in the bottom panel of Fig.\ \ref{fig:BM_shift_specs} we illustrate the Mahan-exciton character of the line shape of the absorption spectrum near the onset, for a high free-electron concentration of 1.1$\times$10$^{19}$ cm$^{-3}$.
To this end, the blue curves show BSE results with (dashed) and without (dot-dashed) the effect of Pauli blocking;
both include free-electron screening of the electron-hole interaction as well as BGR.
Comparing these two curves, shows that Pauli blocking turns the concave line shape (dot-dashed) into a steeper, more convex line shape (dashed) that resembles the case without free electrons much more closely (black dashed).
Hence, this enhancement of the absorption edge can be attributed to the Fermi-edge singularity that only enters when Pauli blocking is included, which is a defining characteristic of the Mahan exciton \cite{Mahan:1967}.
In addition, we also compare to the result that accounts for lattice screening via the dielectric constant (violet curve) and find that in this case the line shape is again more concave.
Thus, Fig.\ \ref{fig:BM_shift_specs} shows that Mahan excitons are the reason that the convex line shape of the case without free electrons is largely preserved up to free-electron concentrations as large as 1.1$\times$10$^{19}$ cm$^{-3}$.
Results that neglect Pauli blocking or approximately capture lattice screening lead to more concave onsets.

\subsection{Exciton binding energy}

\begin{figure}
\includegraphics[width=0.98\columnwidth]{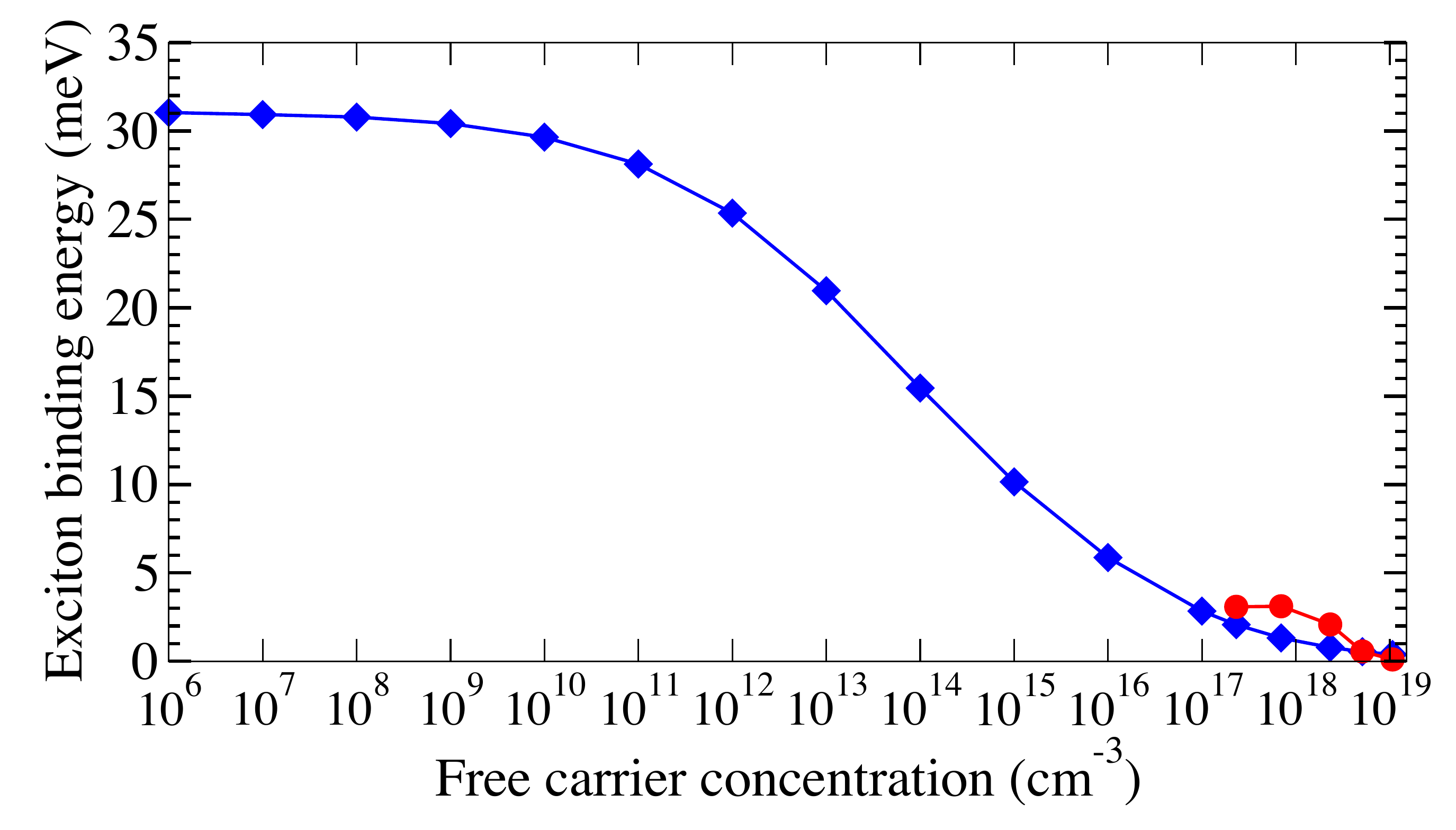}
\caption{\label{fig:exciton_binding_energy}(Color online.)
Exciton binding energy in C MAPbI$_3$ as a function of free-electron concentration, calculated using the BSE$_\mathrm{el+fc}$+$\Delta_{GW_0}$+SOC framework.
Electronic and free-electron screening of the electron-hole interaction are included.
Data with (red) and without (blue) Pauli blocking is compared.
}
\end{figure}

In order to show that the Mahan exciton indeed corresponds to a bound excitonic state that persists in the material despite the presence of free electrons, we computed converged exciton binding energies as the difference between the lowest eigenvalue of the excitonic Hamiltonian and the lowest single-QP excitation energy.
It has been shown before that accurate \textbf{k}-point convergence of the lowest-exciton eigenvalue is challenging and requires dense sampling of the band extrema, in particular for Wannier-Mott type excitons \cite{Fuchs:2008_b}.
We use hybrid \textbf{k}-point meshes to accomplish this and systematically increase the sampling density (see Fig. S1 in the supplemental material).
The densest grid used here samples the entire Brillouin zone by $5\,\times\, 5\,\times\,5$ \textbf{k} points, but the inner third is replaced by a $14\,\times\,14\,\times\,14$ \textbf{k}-point mesh.
The resulting mesh is shifted to center around the direct gap at the $R$-point of the BZ of C MAPbI$_3$.

The resulting value for the exciton-binding energy in C MAPbI$_3$ without free-electrons is $E_\mathrm{b}$=$31.9$ meV.
This is in good agreement with the highest values measured experimentally and other first-principles calculations:
Umari \emph{et al.}\ \cite{Umari:2018} predicted 30 meV and Bokdam \emph{et al.}\ report 45 meV for the tetragonal phase \cite{Kresse_2016}.
The degree of Rashba-Dresselhaus shift is also higher in our work due to a large inversion asymmetry of the relaxed pseudo-cubic phase, leaving fewer states closer to the band-edge.

Next, we compute the change of the exciton binding energy of C MAPbI$_3$ for finite free-electron concentrations in the conduction band, using BSE calculations that account for additional free-electron screening via Eq.\ \eqref{eqn:model_eps_fc}. 
Figure \ref{fig:exciton_binding_energy} shows the resulting decrease of the exciton-binding energy.
For free-electron concentrations around $10^{11}$ cm$^{-3}$, which is comparable to concentrations of charged and shallow defects in highly pure, single-crystalline samples \cite{Shi:2015}, our results show that the exciton binding energy decreased from 31.9 meV to 28.13 meV.
This is still above the thermal dissociation energy at room temperature and, thus, free-electron screening is not a critical factor.
The data in Fig.\ \ref{fig:exciton_binding_energy} also shows a significant drop of the binding energy from 25.35 meV to 10.15 meV for free-electron concentrations of $10^{12}$\,--\,$10^{15}$ cm$^{-3}$.
We note that this is the range where the $q_\mathrm{TF}^2/q^2$ term in Eq.\ \eqref{eqn:model_eps_fc} becomes significant and, thus, free-electron screening becomes the dominant mechanism over electronic interband screening.
We illustrate this explicitly in Fig. S6 of the supplemental material.
This results in the overall decline of the exciton binding energy with increasing free-electron concentration.

At even higher free-electron concentrations between $10^{16}$ and $10^{17}$ cm$^{-3}$, corresponding to those observed in precursor mismatched samples \cite{Wang_2014}, the exciton-binding energy is very small, between 5.87 and 2.84 meV (see Fig.\ \ref{fig:exciton_binding_energy}).
Up to free-electron concentrations of $n_c\approx 2.3\times 10^{17}$ cm$^{-3}$, finite $\mathbf{k}$-point sampling prevents us from explicitly including Pauli blocking in the BSE calculations even for the most dense $\mathbf{k}$-point grid.
Hence, we explore the effect of Pauli blocking due to filling of the conduction band only for higher free-electron concentrations.
For these, Fig.\ \ref{fig:exciton_binding_energy} shows an increase of the binding-energy by up to 2 meV between $n$=$2.3\times 10^{17}$ and 2.3$\times 10^{18}$ cm$^{-3}$, compared to calculations that neglect Pauli blocking. 
This increase has been attributed to the Fermi-edge singularity that arises when Pauli blocking is taken into account and is a characteristic feature of Mahan excitons \cite{Mahan:1967}.
While the small increase of the exciton-binding energy itself is not significant enough to recover a bound exciton at room temperature in samples with a large concentration of free electrons, the Mahan exciton still dominates the line shape of the absorption edge in C MAPbI$_3$, as we discussed above for Fig.\ \ref{fig:BM_shift_specs}.

\section{\label{sec:conclusions}Conclusions and Outlook}

In this work we provide a thorough understanding of the absorption line shape and lowest exciton binding energy of MAPbI$_3$.
Using cutting-edge first-principles theoretical spectroscopy, based on density-functional and many-body perturbation theory, we obtain accurate results for atomic geometries, single-particle electronic structure, and two-particle optical absorption spectra.
These results are a solid foundation for our analysis of free-electron induced effects.
We show that Burstein-Moss shift and band-gap renormalization cancel each other across a large range of free-electron concentrations.
By including these effects as well as free-electron induced dielectric screening when solving the Bethe-Salpeter equation, we explain strongly reduced exciton binding energies, compared to the material without the presence of free electrons.
This elucidates how a wide range of intrinsic free-electron concentrations in MAPbI$_3$ results in a range of exciton binding energies between 2\,--\,30 meV, granting insight into a potential source of variance in experimentally measured exciton binding energies.

Furthermore, we show that the excitons in the presence of free electrons arise from the Fermi edge singularity, proving their Mahan-exciton character.
They determine the line shape of the absorption onset and as a result, the onset still resembles that of the system without free electrons up to very high free-electron concentrations.
Hence, MAPbI$_3$ largely maintains its excellent absorption properties in terms of energy position and line shape.
This can explain why the material remains an excellent photovoltaic absorber even though in real samples free electrons will inevitably be present.
More generally, our results make clear that additional screening of the electron-hole Coulomb interaction by free-electron effects is important in predicting accurate exciton binding energies in MAPbI$_3$, illustrating that a deeper knowledge of electron-hole Coulomb interaction, beyond electronic interband screening, is required.

\begin{acknowledgments}
This work was supported by the National Science Foundation under Grant No.\ CBET-1437230.
This work made use of the Illinois Campus Cluster, a computing resource that is operated by the Illinois Campus Cluster Program (ICCP) in conjunction with the National Center for Supercomputing Applications (NCSA) and which is supported by funds from the University of Illinois at Urbana-Champaign.
This research is part of the Blue Waters sustained-petascale computing project, which is supported by the National Science Foundation (awards OCI-0725070 and ACI-1238993) and the state of Illinois. Blue Waters is a joint effort of the University of Illinois at Urbana-Champaign and its National Center for Supercomputing Applications.
\end{acknowledgments}

\bibliography{./literature.bib}

\newpage

\beginsupplement

\section{Supplemental Information}

\begin{table}
  \caption{\label{tab:lattice}
    Lattice parameters $a$, $b$, $c$ (in \AA) and unit-cell volume $V$ (in \AA$^3$) computed in this work, compared to other theoretical and experimental data.
  }
\begin{tabular}{c|c|c|c}
\hline
  & This Work & Ref. PBEsol\cite{Brivio:2015,Qian:2016} & Exp.\cite{EXP_PARAM_PbI3,Whitfield:2016} \\
\hline
Orthorhombic & & & (13 K) \\
$a$ & 8.365 & 8.350 & 8.571 \\
$b$ & 9.073 & 9.040  & 8.855 \\
$c$ & 12.665 & 12.66 & 12.614 \\
$V$ & 961.27 & 955.63 & 957.35 \\
Tetragonal & & & (180 K) \\
$a$ & 8.711 & 8.700 & 8.970  \\
$b$ & 8.720 & 8.720 & -- \\
$c$ & 12.846 & 12.830 & 12.768 \\
$V$ & 975.810 & 979.78 & 986.510 \\
Cubic & & & 300 K\,--\,350 K \\
$a$ & 6.291 & 6.29 & 6.26-6.31 \\
$b$ & 6.253 & 6.23 & -- \\
$c$ & 6.386 & 6.37 & -- \\
$V$ & 251.12 & 246.62 & 245.31\,--\,251.24 \\
\end{tabular}
\end{table}

\begin{figure}
\includegraphics[width=0.98\columnwidth]{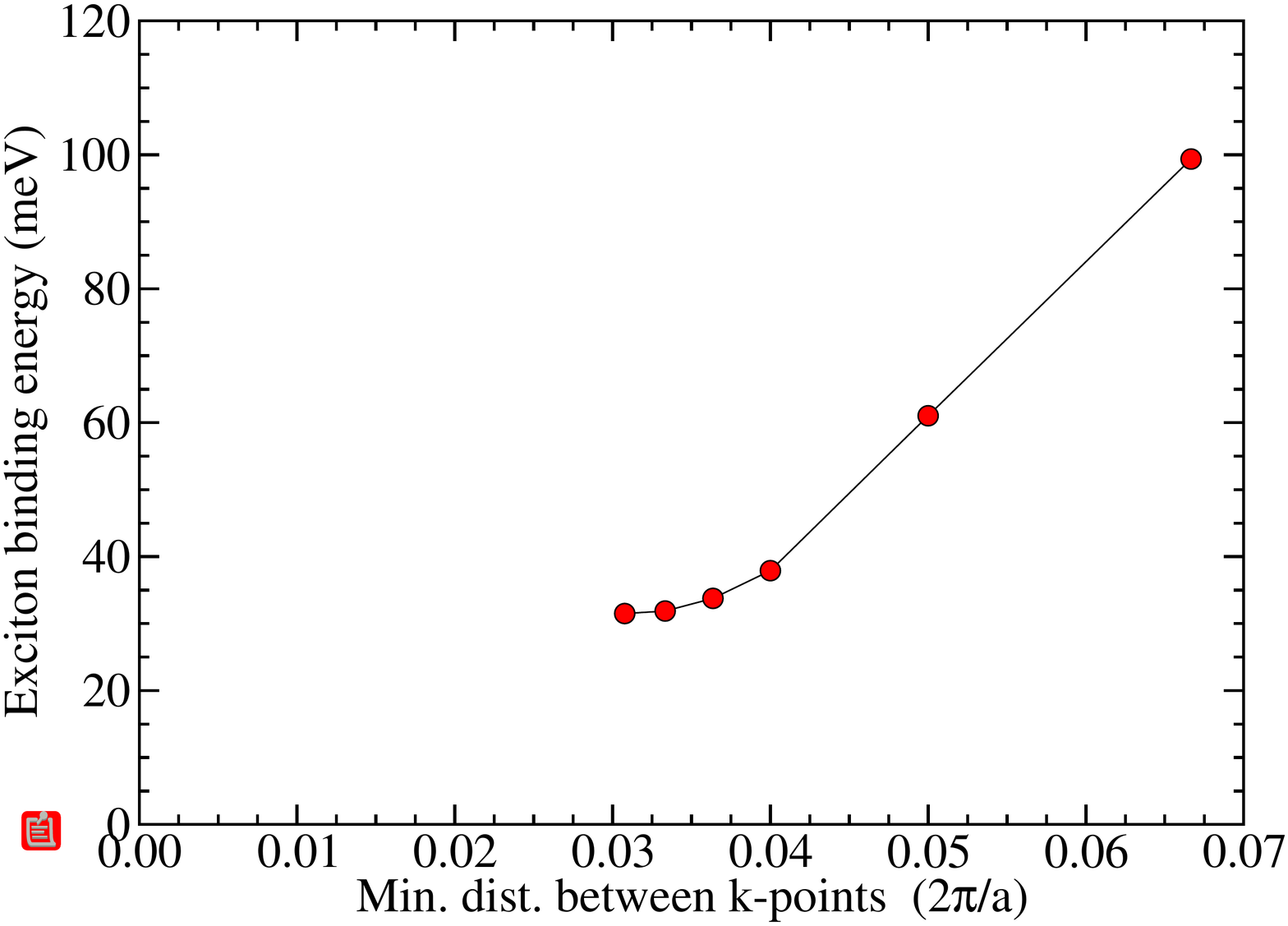}
\caption{\label{fig:Exb_conv}(Color online.)
Convergence of exciton binding energy with respect to $\mathbf{k}$-point grid density. 5:2:15, 5:2:20, 5:20:25, 5:2:27, 5:5:2:30, and 5:2:32.5 type hybrid meshes are used to sample Brillouin zone, ascending in order of $\mathbf{k}$-point sampling density.
}
\end{figure}

\begin{figure*}
\includegraphics[width=0.99\columnwidth]{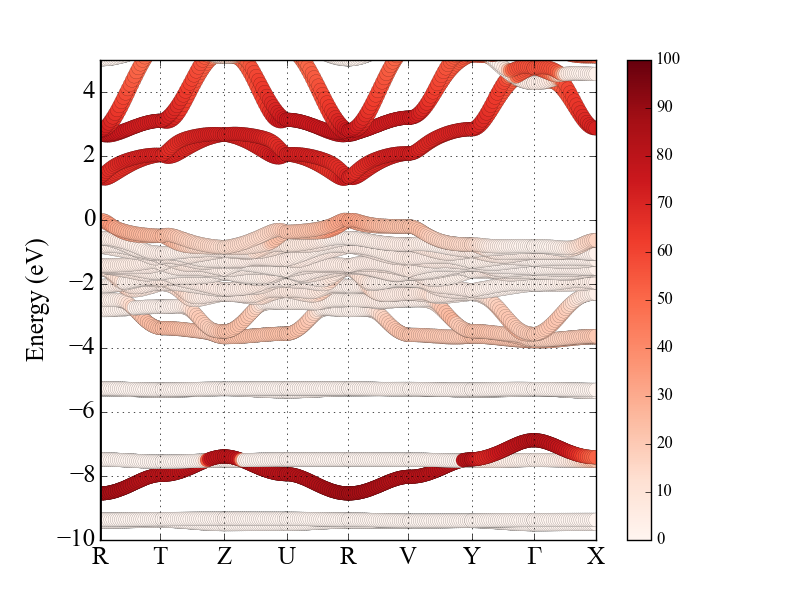}\\
\includegraphics[width=0.99\columnwidth]{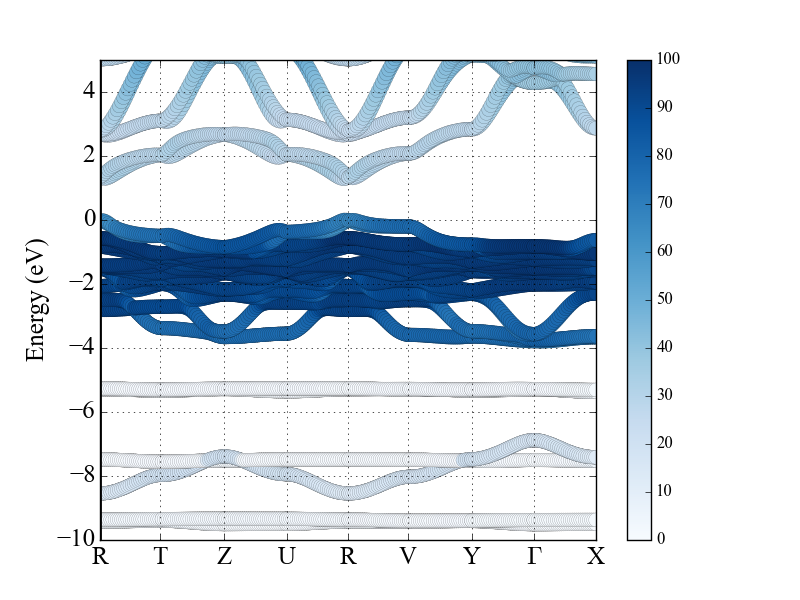}\\
\includegraphics[width=0.99\columnwidth]{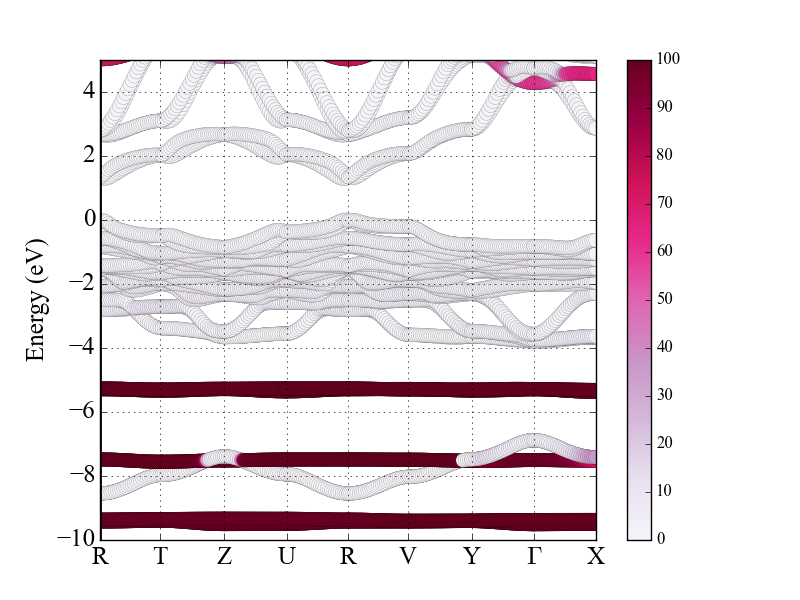}
\caption{\label{fig:atomic_bands_C}(Color online.)
Band structure of C MAPbI$_3$, resolved along atomic orbital contributions of Pb (orange), I (blue), and CH$_3$NH$_3$ (magenta). The color intensity indicates the percent atomic contribution in accordance with the provided color bars. 
}
\end{figure*}

\begin{figure}
\includegraphics[width=0.98\columnwidth]{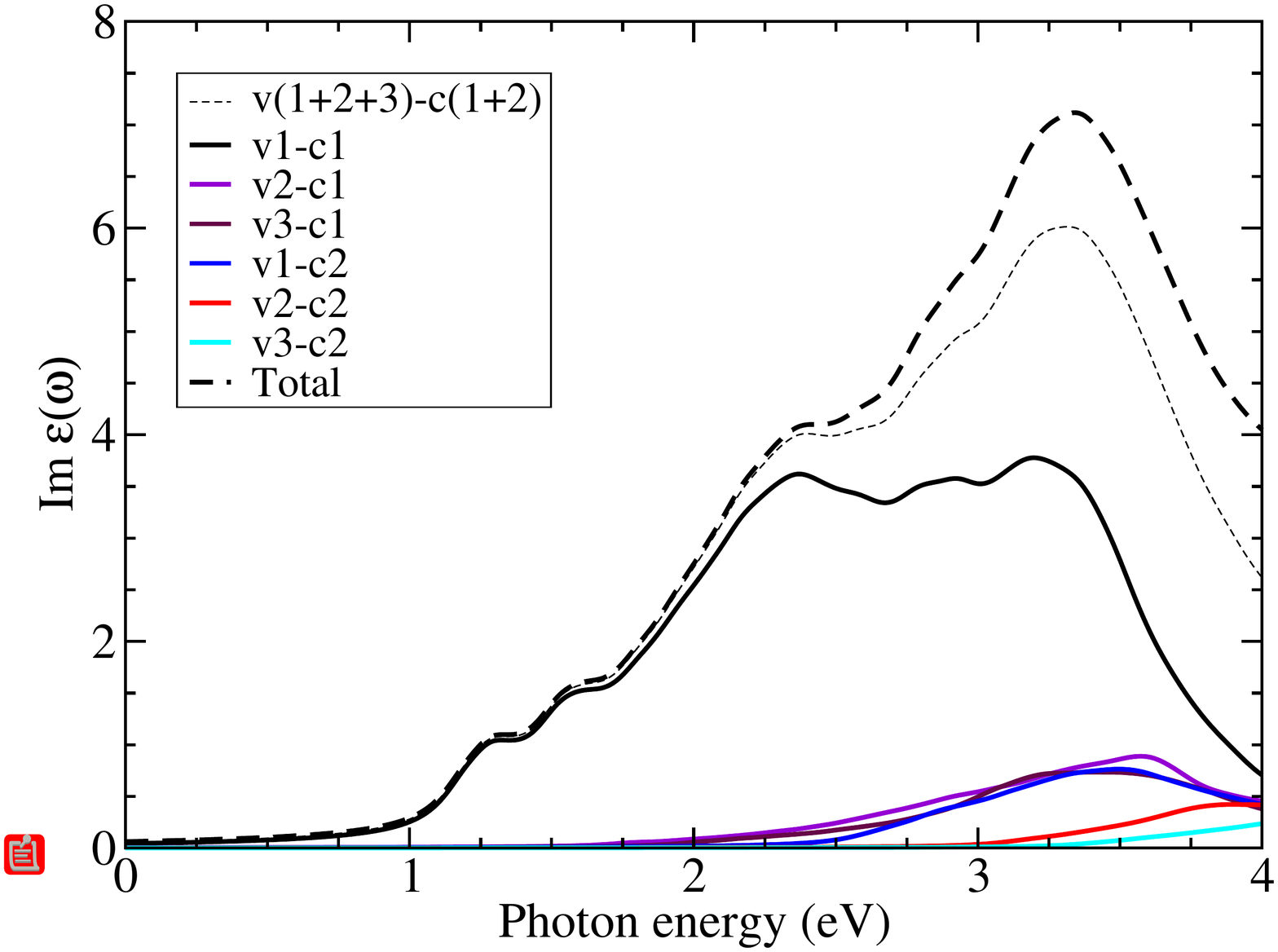}
\caption{\label{fig:band_res_spec}(Color online.)
The imaginary dielectric function of cubic MAPbI$_3$ resolved by band-pair excitation. v indicates valence bands and c indicates conduction bands. Indexing for valence bands starts at the VBM and indices increase as energy decreases. Indexing for conduction bands starts at the CBM and increases in energy. A single index covers a spin-split band pair.}
\end{figure}

\begin{figure}
\includegraphics[width=0.98\columnwidth]{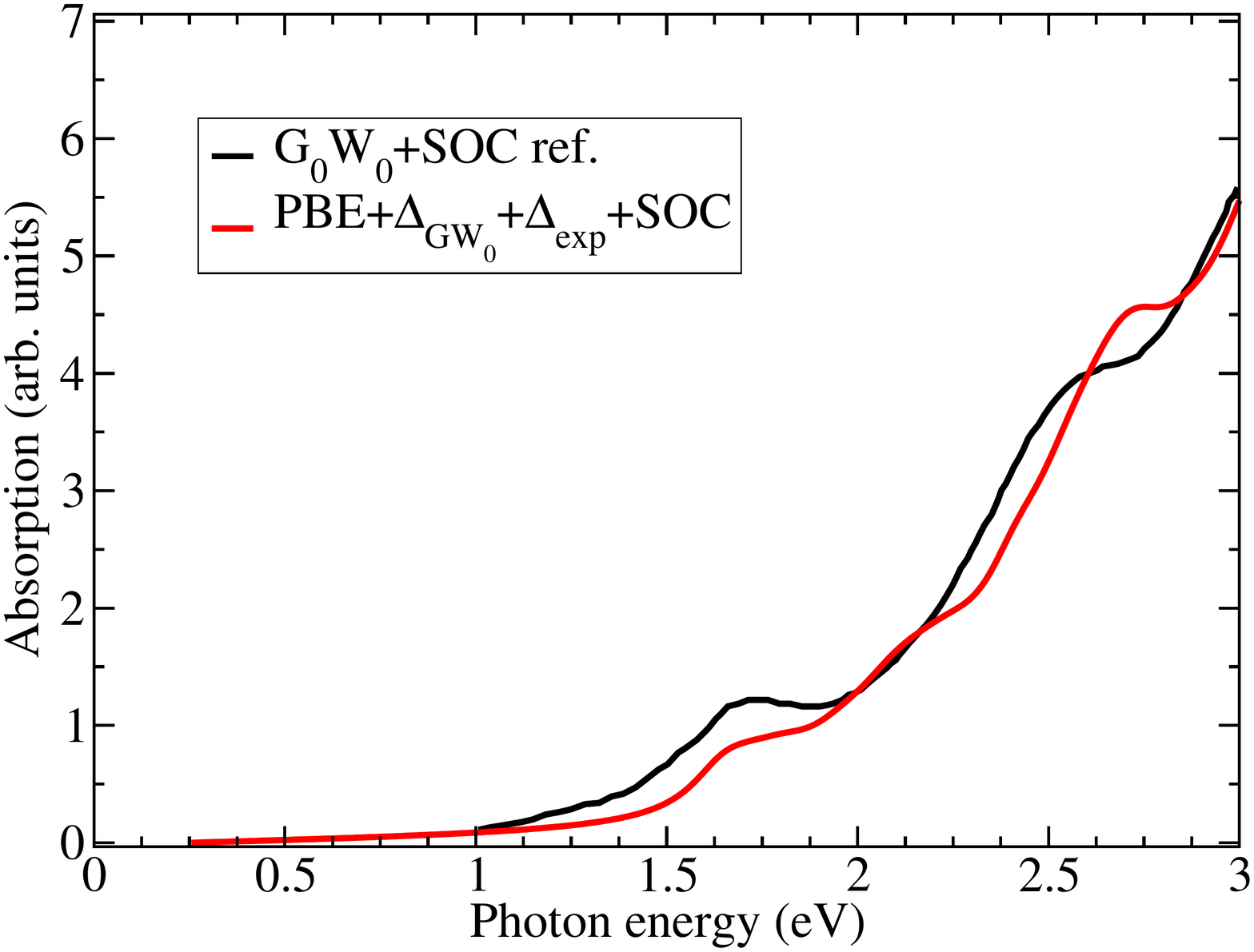}
\caption{\label{fig:Umari_compare}
(Color online.)
Comparison between the optical absorption spectrum of tetragonal MAPbI$_3$ calculated using PBE+$\Delta_{GW_0}$+SOC, shifted further to the experimental gap of 1.6 eV, and 7\,$\times$\,7\,$\times$\,7 $\mathbf{k}$ points with a small random shift (red curve) and the $G_0W_0$+SOC results from Ref.\  \onlinecite{umari_relativistic_2014} (black curve), on a 4\,$\times$\,4\,$\times$\,4 $\mathbf{k}$-point grid.}
\end{figure}

\begin{figure}
\includegraphics[width=0.98\columnwidth]{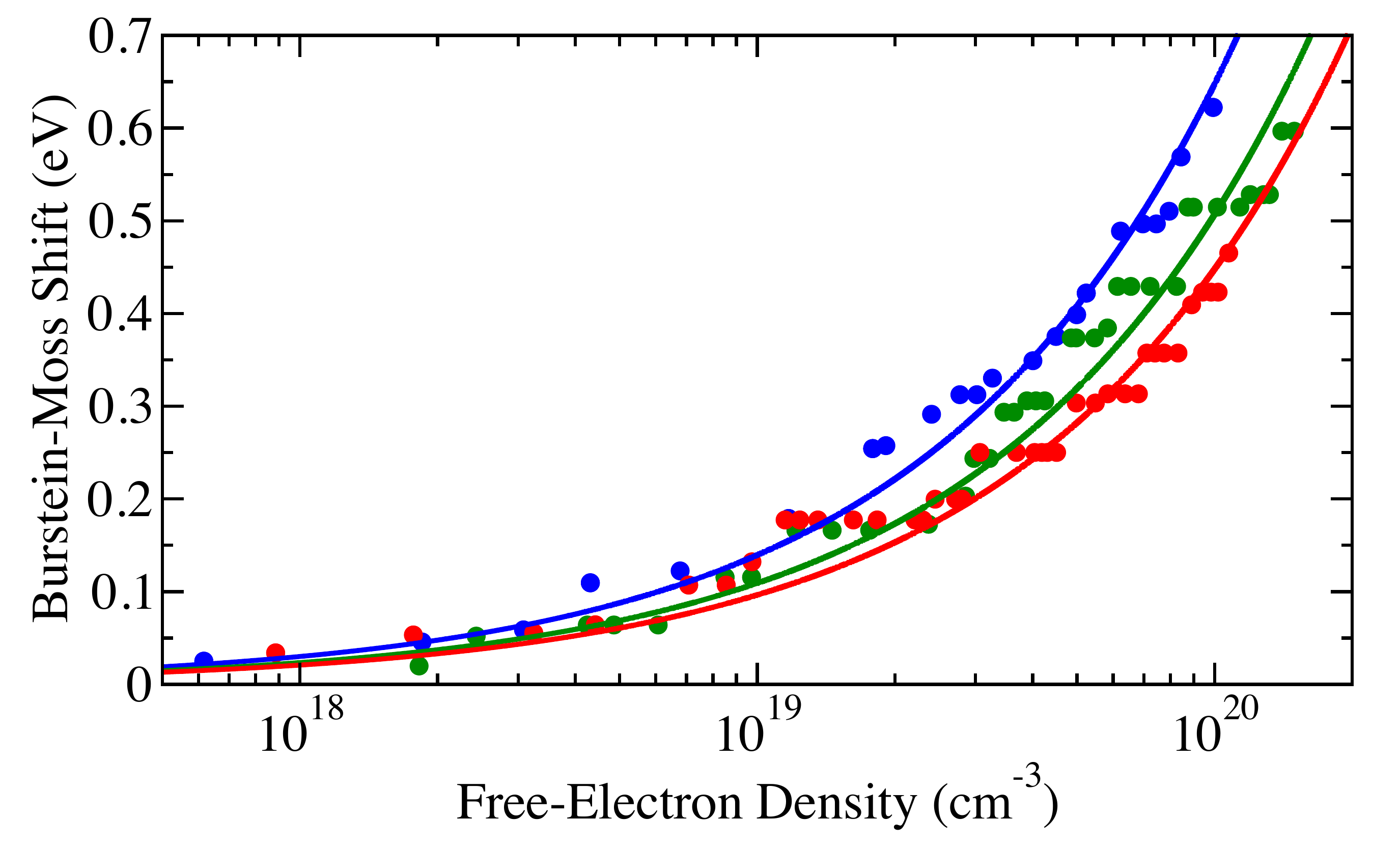}
\caption{\label{fig:bmsbgr}(Color online.)
The calculated Burstein-Moss shifts (circles) of O (blue), T (green) and C (red) as a function of free electron density.
The associated solid lines are functional fits of the form $E_\mathrm{BMS}$=$An_c^{3/2}$.
Dense, hybrid 5:2:15, 5:2:15, and 5:2:32.5 $\mathbf{k}$-point grids were used for O, T, and C phase, respectively (see Ref.\ \onlinecite{Fuchs:2008_b} for nomenclature).
}
\end{figure}

\begin{figure}
\includegraphics[width=0.98\columnwidth]{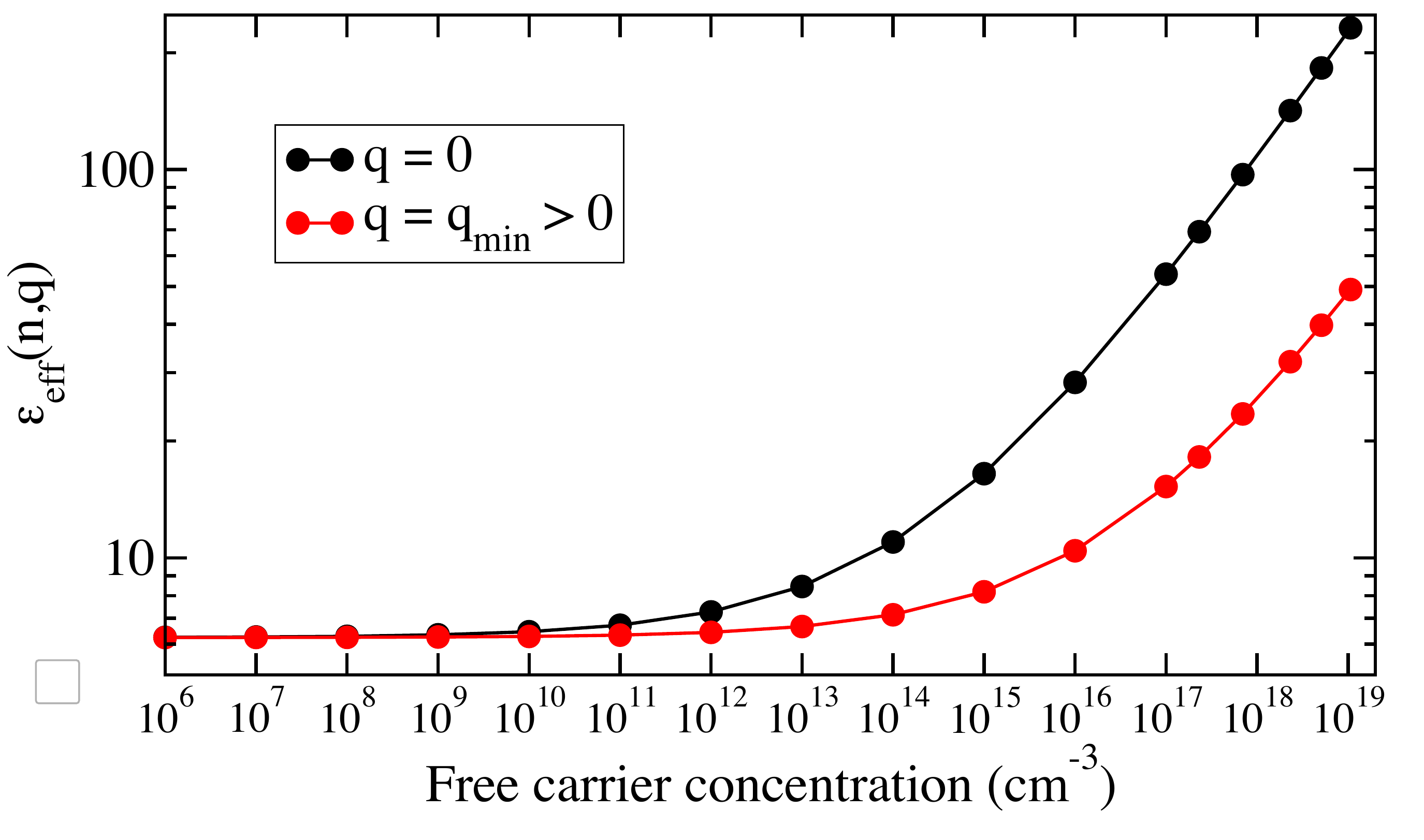}
\caption{\label{fig:eps_eff_qn}(Color online.)
The effective dielectric constant as a function of free-electron concentration and for two wave vectors $q$, resulting from Eqs.\ \eqref{eqn:model_eps_fc} and \eqref{eqn:eps_eff_qn}.
}
\end{figure}

\end{document}